\documentclass[12pt,nofootinbib,preprint]{revtex4}
\pdfoutput=1

\usepackage{amsmath}
\usepackage{amssymb}
\usepackage{anysize}
\usepackage{bold-extra}
\usepackage{xcolor}
\usepackage{enumerate}
\usepackage{fancyhdr}
\usepackage{graphicx}
\usepackage[linktocpage,colorlinks,urlcolor=blue,citecolor=blue,linkcolor=red]{hyperref}
\usepackage{multirow}
\usepackage[final]{pdfpages}
\usepackage{rotating}
\usepackage{textcomp}
\usepackage{url}
\usepackage{verbatim}
\usepackage{wrapfig}
\usepackage{slashed}
\usepackage{subfigure}

\usepackage{amsfonts}
\usepackage{comment}
\usepackage{epigraph}
\usepackage[utf8]{inputenc}
\usepackage{slashed}
\usepackage{bbm}
\usepackage{cancel}
\usepackage{epsf, array, color}
\begin{document}

\preprint{CERN-TH-2020-042, YITP-SB-20-04}

\title{\boldmath Validity of SMEFT studies of VH and VV Production at NLO}

\author{Julien Baglio$^{1}$}
\email{julien.baglio@cern.ch}

\author{Sally Dawson$^{2}$}
\email{dawson@bnl.gov}

\author{Samuel Homiller$^{3}$}
\email{samuel.homiller@stonybrook.edu}

\author{Samuel D. Lane$^{2, 4}$}
\email{samuel.lane@ku.edu}

\author{Ian M. Lewis$^{4}$}
\email{ian.lewis@ku.edu}

\affiliation{$^{1}$CERN, Theoretical Physics Department, CH-1211 Geneva 23, Switzerland\\
$^{2}$Department of Physics, Brookhaven National Laboratory, Upton, New York 11973~ U.S.A.\\
$^{3}$C. N. Yang Institute for Theoretical Physics, Stony Brook University, 
 Stony Brook, New York 11794~ U.S.A.\\
 $^{4}$Department of Physics and Astronomy, University of Kansas, Lawrence, Kansas, 66045~ U.S.A
 }

\begin{abstract}
The production of $W^{\pm}H$, $ZH$, $W^+W^-$, and $W^\pm Z$ pairs
probes non-Standard-Model interactions of quarks, gauge bosons, and the
Higgs boson. New effects can be parameterized in terms of an effective
field theory (EFT) where the Lagrangian is expanded in terms of
higher-dimension operators suppressed by increasing powers of a high
scale $\Lambda$.
We examine the importance of including next-to-leading-order QCD
corrections in global fits to the coefficients of the EFT. The
numerical implications on the fits due to different approaches to
enforcing the validity of the EFT are quantified.
We pay particular attention to the dependence of the fits on the
expansion in $1/\Lambda^2$ since the differences between results
calculated at ${\cal  {O}}(1/\Lambda^2)$ and ${\cal
  {O}}(1/\Lambda^4)$ may give insight into the possible significance
of dimension-8 effects.
\end{abstract}

\maketitle

\section{Introduction}
\label{sec:intro}

One of the most interesting tasks of the high-luminosity phase of the
LHC is to quantify possible experimental differences of Standard Model
(SM) observables from the theoretical predictions.  In the absence of
the discovery of new light particles, effective field theories provide
an efficient means of exploring new physics effects through precision
measurements~\cite{Almeida:2018cld,Biekotter:2018rhp,Grojean:2018dqj,Ellis:2018gqa,Berthier:2016tkq}. Deviations
from the SM can be described in terms of the SM effective field theory
(SMEFT)~\cite{Brivio:2017vri,} which contains an infinite tower of
higher-dimension operators constructed out of SM fields (including an
$SU(2)$ Higgs doublet) that are invariant under the $SU(3)\times
SU(2)\times U(1)$ gauge theory,
\begin{equation}
\mathcal{L}\sim \mathcal{L}_{SM}+\sum_{i,n>4} {C_{i,n}\over \Lambda^{(n-4)}}O_{i,n}\, .
\label{eq:lag}
\end{equation}
The scale $\Lambda$ is taken to generically represent the energy scale
of some unknown UV complete theory and, assuming $\Lambda \gg M_Z$,
the dominant effects typically come from the lowest dimension
operators. In our study, we consider only the dimension-6 operators
and  use the Warsaw operator
basis~\cite{Buchmuller:1985jz,Grzadkowski:2010es}.

Fits to the coefficient functions are done by truncating the
Lagrangian expansion at ${\cal{O}}({1\over \Lambda^2})$. In previous
work, we studied $W^+W^-$ and $W^\pm Z$ production at the LHC in order
to understand the numerical impact of including next-to-leading-order
(NLO) QCD corrections in the fits to the
coefficients~\cite{Baglio:2017bfe,Baglio:2018bkm,Baglio:2019uty}.
Here, we extend the study to include $W^\pm H$ and $ZH$
production~\cite{Alioli:2018ljm} and compute the limits on the
coefficient functions when the cross sections are systematically
expanded to ${\cal{O}}({1\over \Lambda^2})$ and ${\cal{O}}({1\over
  \Lambda^4})$ at leading order (LO) NLO QCD in the SMEFT. We include
anomalous $3$-gauge boson couplings, anomalous gauge boson-Higgs
couplings, and anomalous quark-gauge boson couplings. The SMEFT also
includes interesting 4-point interactions of the form $q {\overline
  q}VH$, ($V=W^\pm,Z$), which lead to novel features. Our work uses
the implementation of these
processes~\cite{Melia:2011tj,Nason:2013ydw,Luisoni:2013kna,Alioli:2018ljm,Baglio:2018bkm,Baglio:2019uty}
in the {\tt POWHEG-BOX} framework~\cite{Frixione:2007vw,Alioli:2010xd}
and we include both $8~\mathrm{TeV}$ and $13~\mathrm{TeV}$ LHC data in
the fits. 

Gauge/Higgs boson pair production has been extensively studied in the
SM. Higher-order SM QCD corrections for $W^+W^-$, $W^\pm Z$, $W^\pm
H$, and $ZH$ exist to
NLO~\cite{Ohnemus:1991kk,Ohnemus:1991gb,Frixione:1992pj,Ohnemus:1994ff,Dixon:1998py,Campbell:1999ah,Campbell:2011bn,Han:1991ia,Ohnemus:1992bd,Baer:1992vx,Stange:1994bb}
and next-to-next-to-leading
order~\cite{Gehrmann:2014fva,Caola:2015rqy,Grazzini:2016swo,Brein:2003wg,Ferrera:2011bk,Ferrera:2014lca,Campbell:2016jau}, 
while electroweak corrections are known to
NLO~\cite{Baglio:2013toa,Bierweiler:2013dja,Billoni:2013aba,Biedermann:2016guo,Biedermann:2017oae,Baglio:2018rcu,Kallweit:2019zez,Denner:2011id} 
for the various processes. The precisely known SM results rely on the
properties of the SM couplings that give cancellations between
Feynman diagrams such that the physical amplitudes do not grow with
energy. Deviations from the SM interactions will spoil these
cancellations~\cite{Hagiwara:1986vm,Hagiwara:1993qt}, potentially
giving measurable effects --- especially in high-energy bins --- and
this property is exploited in the SMEFT fits. Higher-order QCD corrections,
including effects of the anomalous triple-gauge-boson couplings, exist
at NLO for diboson
production~\cite{Baur:1994aj,Baur:1995uv,Dixon:1999di,Chiesa:2018lcs}
and have been extended to include also the effect of anomalous quark
couplings~\cite{Baglio:2017bfe,Baglio:2018bkm,Baglio:2019uty}. $W^\pm
H$ and $Z H$ channels are also known at NLO QCD including SMEFT
operators~\cite{Campanario:2014lza,Granata:2017iod,Alioli:2018ljm}.
 
In this work, we perform a  fit to the dimension-6 coefficients
relevant for the $W^+ W^-$, $W^\pm Z$, $W^\pm H$, and $ZH$ channels at
NLO QCD. At NLO, the additional jet reduces the sensitivity to
anomalous couplings and this  effect is often compensated for by
imposing a jet veto above some $p_T$. 
Our focus is on understanding the numerical importance of the NLO
SMEFT QCD corrections and  the jet veto cuts on the sensitivity to the
SMEFT coefficients~\cite{Azatov:2019xxn,Campanario:2014lza}.

Since we are considering dimension-6 operators, the Lagrangian of
Eq. \ref{eq:lag} generates terms of
$\displaystyle{\cal{O}}\left(({\textrm{Energy}
    \over\Lambda})^2\right)$.  If there is some generic coupling
strength, $g_{\textrm{EFT}}$, associated with the EFT, there are also
terms of
$\displaystyle{\cal{O}}\left(({g_{\textrm{EFT}}v\over\Lambda})^2\right)$. In
order for a weak-coupling EFT expansion to be valid, both classes of
terms must be small. We study the regions in our fits where these
criteria are satisfied~\cite{Contino:2016jqw}.
We further study the numerical effects of including $1/\Lambda^2$ or
$1/\Lambda^4$ contributions. It has been suggested that the difference
between results obtained at $1/\Lambda^2$ or $1/\Lambda^4$ could be an
indication of the size of the dimension-8 contributions, which are
also formally of
${\cal{O}}(1/\Lambda^4)$~\cite{Hays:2018zze,Alte:2018xgc}.

In Section~\ref{sec:basics}, we review the details of the SMEFT that
are relevant for our study and the implementation in the {\tt
  POWHEG-BOX} framework. Section~\ref{sec:powheg} demonstrates the
effects of NLO corrections on distributions, and the effects of jet
veto cuts on the sensitivity of these distributions to anomalous
couplings. Finally, section~\ref{sec:res} presents the results of both
profiled and projected fits, while quantifying the effects of the NLO
corrections, the effects of order $1/\Lambda^4$ on the fits, and a
discussion of the applicability of our fits in the
context of a weakly-coupled theory.

\section{Basics}
\label{sec:basics}

The production rates for $W^+W^-,~W^\pm Z, ~W^\pm H$, and $ZH$ at high
energy are extremely sensitive to new-physics
effects~\cite{Falkowski:2015jaa,Franceschini:2017xkh,Liu:2019vid,Grojean:2018dqj}. We
parameterize  possible new interactions in terms of general
CP-conserving, Lorentz-invariant interactions, neglecting dipole
interactions since they do not interfere with the SM results for these
processes. We also neglect flavor effects.
The correspondence between various SMEFT basis choices is
straightforward~\cite{Falkowski:2015wza}, and we will always use the
Warsaw basis for which the Feynman rules  and operator definitions can
be obtained from~\cite{Dedes:2017zog,Brivio:2017bnu}.

In the Warsaw basis, the relationships between inputs are altered from
those of the SM.  Taking  the measured values of $G_F$, $M_W$, and
$M_Z$ as inputs, the tree-level shifts in the couplings
are~\cite{Brivio:2017bnu},
\begin{eqnarray}
{\delta G_F\over G_F} &=& {v^2\over \Lambda^2}\biggl\{ C_{Hl}^{(3)}-{1\over 2} C_{ll}\biggr\}\, ,
\nonumber \\
{\delta M_Z^2\over M_Z^2} &=&
{v^2\over 2\Lambda^2}\biggl\{ C_{HD}+{4M_W\over M_Z}\sqrt{1-{M_W^2\over M_Z^2}}C_{HWB}\biggr\}\, ,
\nonumber \\
{\delta M_W^2\over M_W^2} &=& 0\, ,
\nonumber \\
\delta g_Z &=& -\frac{v^2}{\Lambda^2}\left(\delta v +\frac14 C_{HD}^{}\right) \, ,
\nonumber \\
\delta v  &=& C_{Hl}^{(3)} - \frac12 C_{ll}^{}\, ,
\nonumber \\
\delta s_W^2 &=&  -\frac{v^2}{\Lambda^2} \frac{s_W^{} c_W^{}}{c_W^2-s_W^2}\left[2 s_W^{} c_W^{}\left(\delta v + \frac14 C_{HD}^{}\right) + C_{HWB}^{}\right] \, , \nonumber
\label{eq:pardefs}
\end{eqnarray}
where we write the SMEFT quantity $x$ in terms of the measured value ${\hat{x}}$ and the
shift $\delta x$: $x={\hat{x}}-\delta x$. It should be noted that
$\delta v$ does not follow this method. Instead it is the
dimensionless shift to $G_F$ coming from  muon decay.
With these inputs, $g^2=4\sqrt{2}G_F M_W^2$, $\cos\theta_W\equiv
c_W=M_W/M_Z$, and $e=g \sin\theta_W\equiv g s_W$.  In our fits we will
take $C_{Hl}^{(3)} = \frac12 C_{ll}^{}$ = 0, since these parameters
are tightly constrained by muon decays~\cite{Jenkins:2017jig}. 

Historically, the SMEFT interactions have been studied from a general
interaction perspective.  The $3$-gauge boson vertices can be written
as,
\begin{eqnarray}
 \mathcal{L}_{WWZ}&=&
-ig_{WWZ}\biggl[g_1^Z\left(W^+_{\mu\nu}W^{-\mu}Z^\nu-W_{\mu\nu}^-W^{+\mu}Z^\nu\right)+\kappa^ZW^+_\mu
            W^-_\nu Z^{\mu\nu}+\frac{\lambda^Z}{M^2_W}W^+_{\rho\mu}{W^{-\mu}}_\nu Z^{\nu\rho}\biggr], 
\nonumber \\
\mathcal{L}_{WW\gamma}&=&-            ig_{WW\gamma}\biggl[\left(W^+_{\mu\nu}W^{-\mu}\gamma^\nu-W_{\mu\nu}^-W^{+\mu}\gamma^\nu\right)+\kappa^\gamma W^+_\mu
            W^-_\nu \gamma^{\mu\nu}+\frac{\lambda^\gamma}{M^2_W}W^+_{\rho\mu}{W^{-\mu}}_\nu \gamma^{\nu\rho}\biggr],
\label{eq:lagdef}
\end{eqnarray}  
with $g_{WW\gamma}=e$, $g_{WWZ}=g \cos\theta_W$, $g_1^Z=1+\delta
g_1^Z$, and $\kappa^{Z,\gamma}=1+\delta \kappa^{Z,\gamma}$.  $SU(2)$
gauge invariance implies
\begin{eqnarray}
\delta g_1^Z&=&\delta \kappa^Z+{s_W^2\over c_W^2}\delta \kappa^\gamma
                \, , \nonumber \\
\lambda^\gamma&=&\lambda^Z\, .
\end{eqnarray}
Expressions for the anomalous gauge couplings in the Warsaw basis are
given in Table~\ref{tab:rgb}~\cite{Baglio:2017bfe,Zhang:2016zsp,Berthier:2015oma,Dedes:2017zog}. 

Neglecting dipole interactions, the quark-gauge boson couplings can be
written as,
\begin{eqnarray}
  \mathcal{L}_{ffV}&\equiv &g_ZZ_\mu\biggl[g_L^{Zq}+\delta g_{L}^{Zq}\biggr]
  {\overline q}_L\gamma_\mu q_L\
 +g_ZZ_\mu\biggl[g_R^{Zq}+\delta g_{R}^{Zq}\biggr]
  {\overline q}_R\gamma_\mu q_R\nonumber \\
  &&+{g\over \sqrt{2}}\biggl\{W_\mu\biggl[(1+\delta g_{L}^W){\overline u}_L\gamma_\mu d_L
  +\delta g_R^W
  {\overline u}_R\gamma_\mu d_R\biggr] +h.c.\biggr\}\, ,
  \label{eq:dgdef}
 \end{eqnarray}
 with
$g_Z=e/(c_W^{}s_W^{})= g/c_W$.  The SM quark interactions  are:
\begin{eqnarray}
g_R^{Zq}&=&-s_W^2 Q_q\quad{\rm and}\quad g_L^{Zq}=T_3^q -s_W^2 Q_q,
\end{eqnarray}
where $T_3^q=\pm \displaystyle \frac{1}{2}$ and $Q_q$ is the electric
charge. Expressions for the  anomalous fermion- gauge couplings in the
Warsaw basis are given in Table~\ref{tab:ferm}~\cite{Baglio:2017bfe,Zhang:2016zsp,Berthier:2015oma,Dedes:2017zog}.
 
Finally, the relevant Higgs couplings  (again neglecting dipole
interactions) are described by,
\begin{eqnarray}
\mathcal{L}_{VVH}&=&\mathcal{L}_{H}^{SM} +c_{1Z}HZ_\mu Z^\mu +c_{2Z}Z_{\mu\nu}Z^\mu\partial^\nu H+c_{3Z}HZ_{\mu\nu}Z^{\mu\nu}\nonumber \\
&&+c_{1W}HW_\mu^+W^{-\mu}+c_{2W}\biggl(W_{\mu\nu}^+W^{-\mu}+W_{\mu\nu}^-W^{+\mu}\biggr)\partial^\nu H+c_{3W}HW_{\mu\nu}W^{\mu\nu}
\nonumber \\
&&+d_{1Z}^{Lf}({\overline{ f}}_{L}\gamma_\mu f_{L})Z^\mu H
+d_{1Z}^{Rf}({\overline{ f}}_{R}\gamma_\mu f_{R})Z^\mu H
\nonumber \\ && +\biggl\{
d_{1W}^L({\overline{ u}}_{L}\gamma_\mu d_{L})W^\mu H
+d_{1W}^R({\overline{ u}}_{R}\gamma_\mu d_{R})W^\mu H + h.c.
\biggr\}
\, ,
\label{eq:ffh}
\end{eqnarray}
where $\mathcal{L}_{H}^{SM}$ contains the relevant SM Higgs interactions.
In the Warsaw basis, the effects of $c_{2W}$ and $c_{2Z}$ are
eliminated using the equations of motion.
Expressions for the anomalous Higgs couplings are given in the Warsaw
basis in Table~\ref{tab:ghh}~\cite{Dedes:2017zog}.
Finally, the SMEFT contains two $4$-point operators that contribute to
$VH$ production, $O_{Hq}^{(1)}$ and
$O_{Hq}^{(3)}$~\cite{Dedes:2017zog}.
We note that the parameterizations of
Eqs.~\ref{eq:lagdef}-\ref{eq:ffh} are closely related to those of the
Higgs basis~\cite{Gupta:2014rxa,Falkowski:2015fla}.
Finally, we assume that the $Hb{\overline{b}}$ coupling is SM-like,
since we expect the anomalous coefficients involving the $b$ and
the Higgs to be suppressed by factors of $m_b \over v$ compared to the
effects of  other operators.

We are now ready to count the parameters appearing in our study. 
The $W^+W^-$ and $WZ$ processes can be described by $7$ independent
couplings which we take to be,
\begin{equation}
\delta g_1^Z,\ \delta \kappa_Z,\  \delta \lambda_Z,\  \delta g_L^{Zu},\  \delta g_L^{Zd},\  \delta g_R^{Zu},\  \delta g_R^{Zd} \, .
\end{equation}
Neglecting possible right-handed $W$ couplings (since they are known
to be small~\cite{Tanabashi:2018oca}), the $W^\pm H $ process depends
on $3$ combinations of couplings,
\begin{equation}
\biggl(C_{1W}, C_{3W}\biggr),\  \delta g_L^{W}=\delta g_L^{Zu}-\delta g_L^{Zd},\  C_{Hq}^{(3)}\, ,
\end{equation}
where by $\biggl(C_{1W}, C_{3W}\biggr)$ we mean the combination of
these coefficients that comes into the $WWH$ vertex.
$ZH$  production  is sensitive to
\begin{eqnarray}
\biggl(C_{1Z}, C_{ZZ}, C_{3Z}\biggr) ,\ \delta g_L^{Zu},\ \delta
  g_L^{Zd},\ \delta g_R^{Zu},\ \delta g_R^{Zd} ,\
  \biggl(C_{Hq}^{(3)},\ C_{Hq}^{(3)} \biggr) \, ,
\end{eqnarray}
where $\biggl(C_{1Z}, C_{ZZ}, C_{3Z}\biggr)$ and $\biggl(C_{Hq}^{(3)},
C_{Hq}^{(3)}\biggr)$ are the combination of coefficients that affect
ZH production.
Since we fit to $W^+W^-,W^\pm Z,W^\pm H,$ and $WZ$ there are 10
relevant parameters that we express in terms of their Warsaw basis
coefficients. We note that the purpose of our study is not to do a
complete global fit, but to quantify the effects of the QCD
corrections and the expansion in powers of $1/\Lambda^2$ on fits to
these observables.

\begin{table}
\centering
\begin{tabular}{|c||c|}
\hline
& Warsaw  Basis 
\\
\hline\hline
$\delta g_1^Z$ & $\frac{v^2}{\Lambda^2}\frac{1}{c_W^2-s_W^2}\left(\frac{s_W^{}}{c_W^{}}C_{HWB}^{} + \frac14 C_{HD}^{} +\delta v\right)$\\\hline
$\delta \kappa_{}^Z$ &$\frac{v^2}{\Lambda^2}\frac{1}{c_W^2-s_W^2}\left(2 s_W c_W C_{HWB}^{} + \frac14 C_{HD}^{} +\delta v\right)$ \\ \hline
$\delta \kappa_{}^\gamma$ & $-\frac{v^2}{\Lambda^2}\frac{c_W^{}}{s_W^{}}C_{HWB}^{}$  \\ \hline
$\lambda_{}^\gamma$ &$\frac{v}{\Lambda^2} 3 M_W^{} C_{W}$  \\ \hline
$\lambda_{}^Z$ & $ \frac{v}{\Lambda^2} 3 M_W^{} C_{W}$ \\ \hline
\hline
\end{tabular}
\caption{Anomalous 3-gauge-boson couplings in the
  Warsaw basis.~$\delta v$ is given in Eq. \ref{eq:pardefs}.}
\label{tab:rgb}
\end{table}

\begin{table}
\centering
\begin{tabular}{|c||c|}
\hline
& Warsaw  Basis \\
\hline\hline
$\delta g_L^{Zu}$&$ -\frac{v^2}{2\Lambda^2}\left(C_{Hq}^{(1)}-C_{Hq}^{(3)}\right) + \frac12 \delta g_Z + \frac23\left(\delta s_W^2 - s_W^2 \delta g_Z^{}\right)$ \\
\hline
$\delta g_L^{Zd}$& $-\frac{v^2}{2\Lambda^2}\left(C_{Hq}^{(1)}+C_{Hq}^{(3)}\right) - \frac12 \delta g_Z - \frac13\left(\delta s_W^2 - s_W^2 \delta g_Z^{}\right)$\\
\hline
$\delta g_R^{Zu}$&$-\frac{v^2}{2\Lambda^2} C_{Hu} + \frac23\left(\delta s_W^2 - s_W^2\delta g_Z^{}\right)$ \\
\hline
$\delta g_R^{Zd}$& $-\frac{v^2}{2\Lambda^2} C_{Hd} - \frac13\left(\delta s_W^2 - s_W^2\delta g_Z^{}\right)$\\
\hline
$\delta g_L^W$&  $ \frac{v^2}{\Lambda^2}C_{Hq}^{(3)} + c_W^2\delta g_Z^{} + \delta s_W^2$\\
\hline
\hline
\end{tabular}
\caption{Anomalous fermion couplings in the Warsaw  basis.}
\label{tab:ferm}
\end{table}

\begin{table}
\centering
\begin{tabular}{|c||c|}
\hline
& Warsaw  Basis 
\\
\hline\hline
$c_{1W}$ &  $2M_W^2\sqrt{G_F\sqrt{2}}\biggl\{
{v^2\over \Lambda^2}\biggr(C_{H\square}-{1\over 4}C_{HD}\biggr)+{\delta M_W^2\over M_W^2}
+{\delta G_F\over  2 G_F}\biggr\}$\\
\hline
$c_{1Z}$ &  $2M_Z^2\sqrt{G_F\sqrt{2}}\biggl\{
{v^2\over \Lambda^2}\biggr(C_{H\square}+{3\over 8}C_{HD}+s_Wc_WC_{HWB} \biggr)+{\delta M_Z^2\over M_Z^2}
+{\delta G_F\over  2 G_F}\biggr\}$ \\
\hline
$c_{3W}$ & ${vC_{HW}\over \Lambda^2}$ \\
\hline
$c_{3Z}$ &  ${v\over \Lambda^2}\biggl(c_W^2C_{HW}+s_W^2C_{HB}+s_Wc_W C_{HWB}\biggr)$\\
\hline
$d_{1Z}^{Ru}$  &  ${2M_Z\over\Lambda^2}C_{Hu}$
\\
\hline
$d_{1Z}^{Lu}$  & ${2M_Z\over \Lambda^2} \biggl(C_{Hq}^{(1)}-C_{Hq}^{(3)}\biggr)$
\\
\hline
$d_{1Z}^{Rd}$ & ${2M_Z\over\Lambda^2}C_{Hd}$\\
$d_{1Z}^{Ld}$ &${2M_Z\over \Lambda^2} \biggl(C_{Hq}^{(1)}+C_{Hq}^{(3)}\biggr)$ 
\\
\hline
$d_{1W}^{R}$   & $-\sqrt{2}{M_W\over \Lambda^2} C_{Hud}$\\
\hline
$d_{1W}^{L}$  & $-2\sqrt{2}{M_W\over \Lambda^2} C_{Hq}^{(3)}$\\
\hline\hline
\end{tabular}
\caption{Anomalous Higgs gauge boson couplings in the
  Warsaw  basis .}
\label{tab:ghh}
\end{table}

\section{Results}\label{sec:powheg}

\subsection{Simulation}
\label{sec:simul}

For each process  ($W^+W^-, W^\pm Z, W^\pm H$, and $ZH$), we introduce
anomalous couplings in the Warsaw basis and utilize existing
implementations in the {\tt POWHEG-BOX} framework,  working to NLO QCD
within the SMEFT\footnote{This public tool can  be  found at
  \href{http://powhegbox.mib.infn.it}{http://powhegbox.mib.infn.it}. 
We make use of the \texttt{WWanomal}, \texttt{WZanomal},
\texttt{HW\_smeft} and \texttt{HZ\_smeft} user processes introduced in
previous works~\cite{Baglio:2018bkm, Baglio:2019uty, Alioli:2017ces,
  Alioli:2018ljm}}. 
We consider only the leptonic decays of the gauge bosons and the Higgs
decay to $b {\overline b}$. 
Using the {\tt POWHEG-BOX-V2} program, we compute primitive
differential cross sections that allow us to scan over anomalous
couplings in an efficient manner~\cite{Baglio:2017bfe}. The primitive
cross sections  are extracted in such a way as  to allow for  the
consistent calculation at either linear,
$\mathcal{O}(\frac{1}{\Lambda^2})$,  or quadratic,
$\mathcal{O}(\frac{1}{\Lambda^4})$, order. The results shown in the
following sections use CTEQ14qed PDFs and we fix the
renormalization/factorization scales to $M_Z/2$ .

\subsection{Distributions in the Presence of Radiation}
\label{sec:dists}

A principal advantage of the SMEFT framework is that it allows for a
systematic study of distributions in the presence of new physics
modifying the couplings between  the SM fields.
An important goal is thus to understand how to extract the maximum
possible amount of information from these distributions.
In this light, it is crucial to understand how these distributions are
influenced by higher-order corrections, particularly in the presence
of extra QCD radiation.
The presence of additional jets can substantially change the
distributions, washing out effects present at tree level, and in some
cases, dramatically change the results of a fit to experimental
data~\cite{Baglio:2019uty,Baglio:2018bkm,Baglio:2017bfe}.
The effects of a jet veto have been studied in the past by considering
extra partons at the matrix element level at leading
order~\cite{Azatov:2019xxn,Franceschini:2017xkh} and at NLO QCD  in
the SM~\cite{Campanario:2014lza}.
Our study includes the full NLO QCD SMEFT corrections and clearly
demonstrates the difference between including $1/\Lambda^2$ terms and
$1/\Lambda^4$  contributions in the cross sections.

\begin{figure}
  \includegraphics[width=0.49\linewidth]{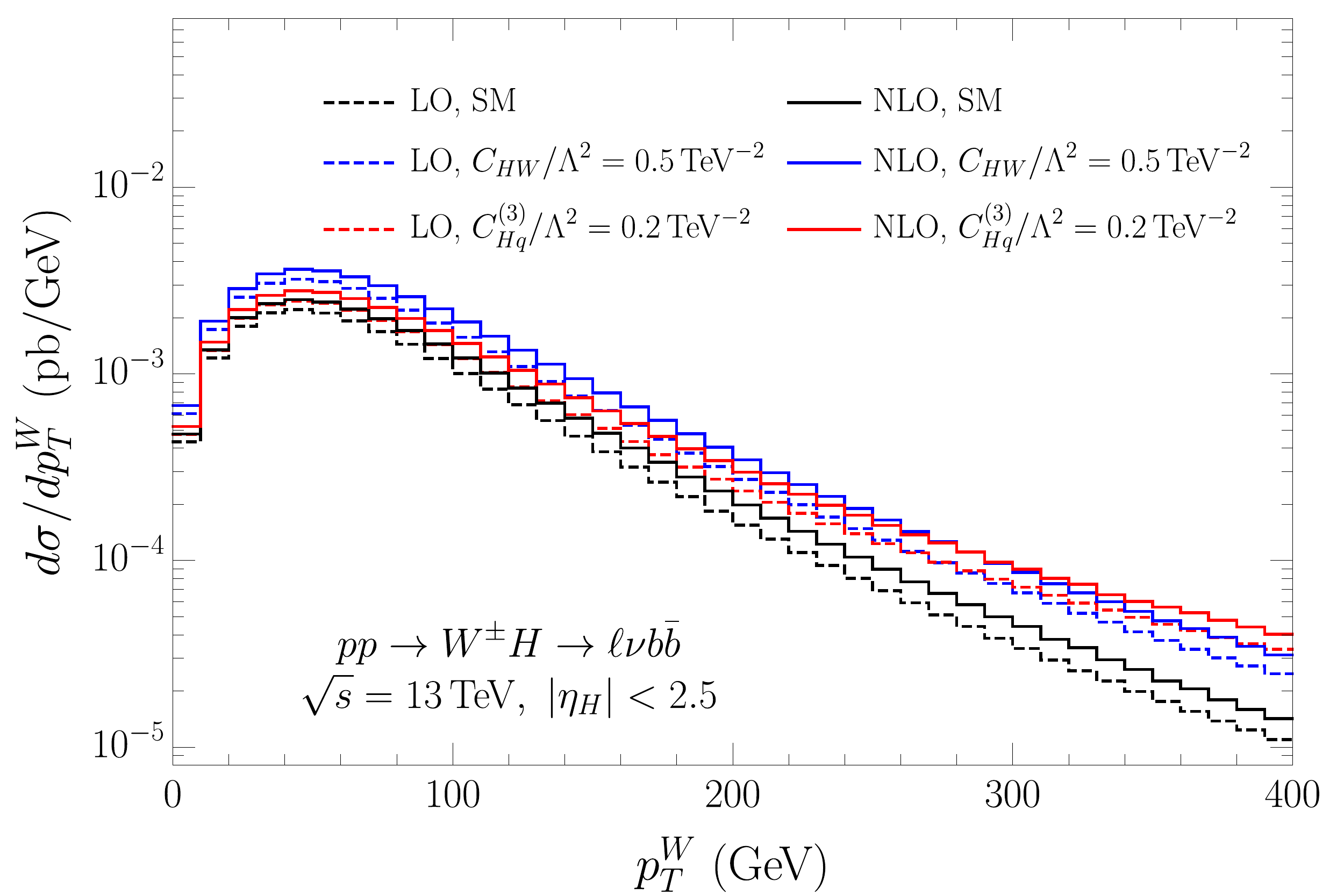}
  \hfill
  \includegraphics[width=0.49\linewidth]{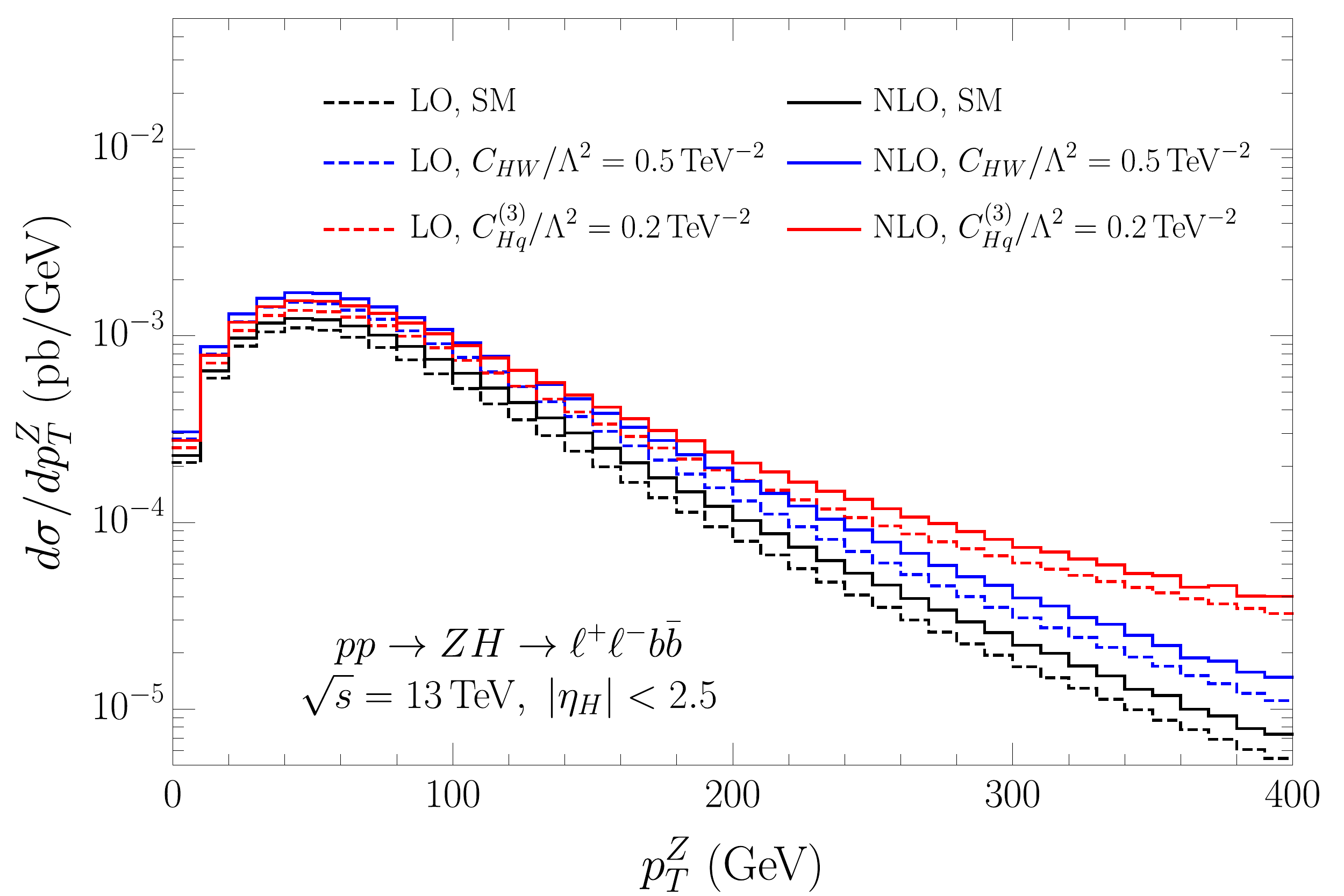}\\
  \vskip 0.2cm
  \includegraphics[width=0.49\linewidth]{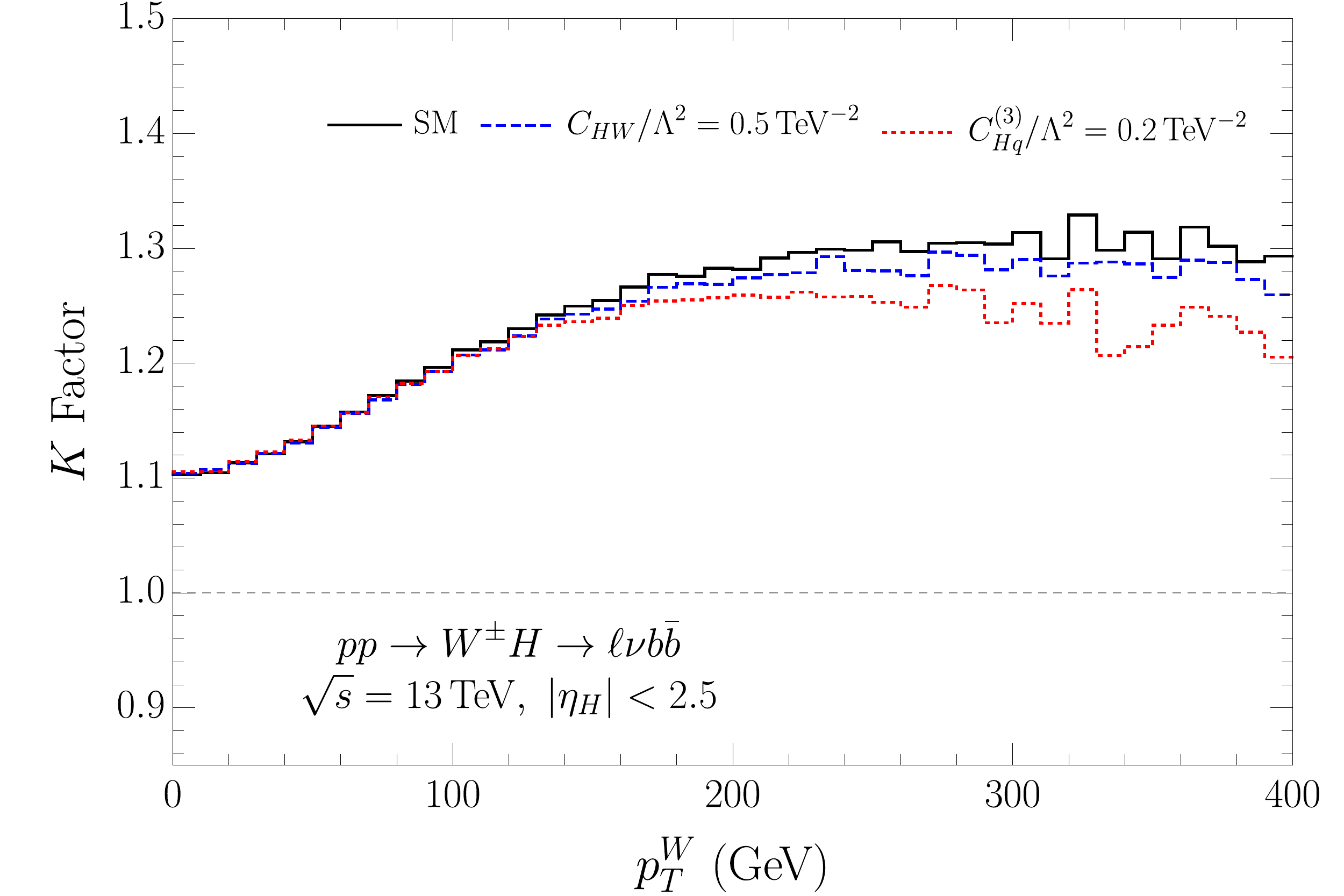}
  \hfill
  \includegraphics[width=0.49\linewidth]{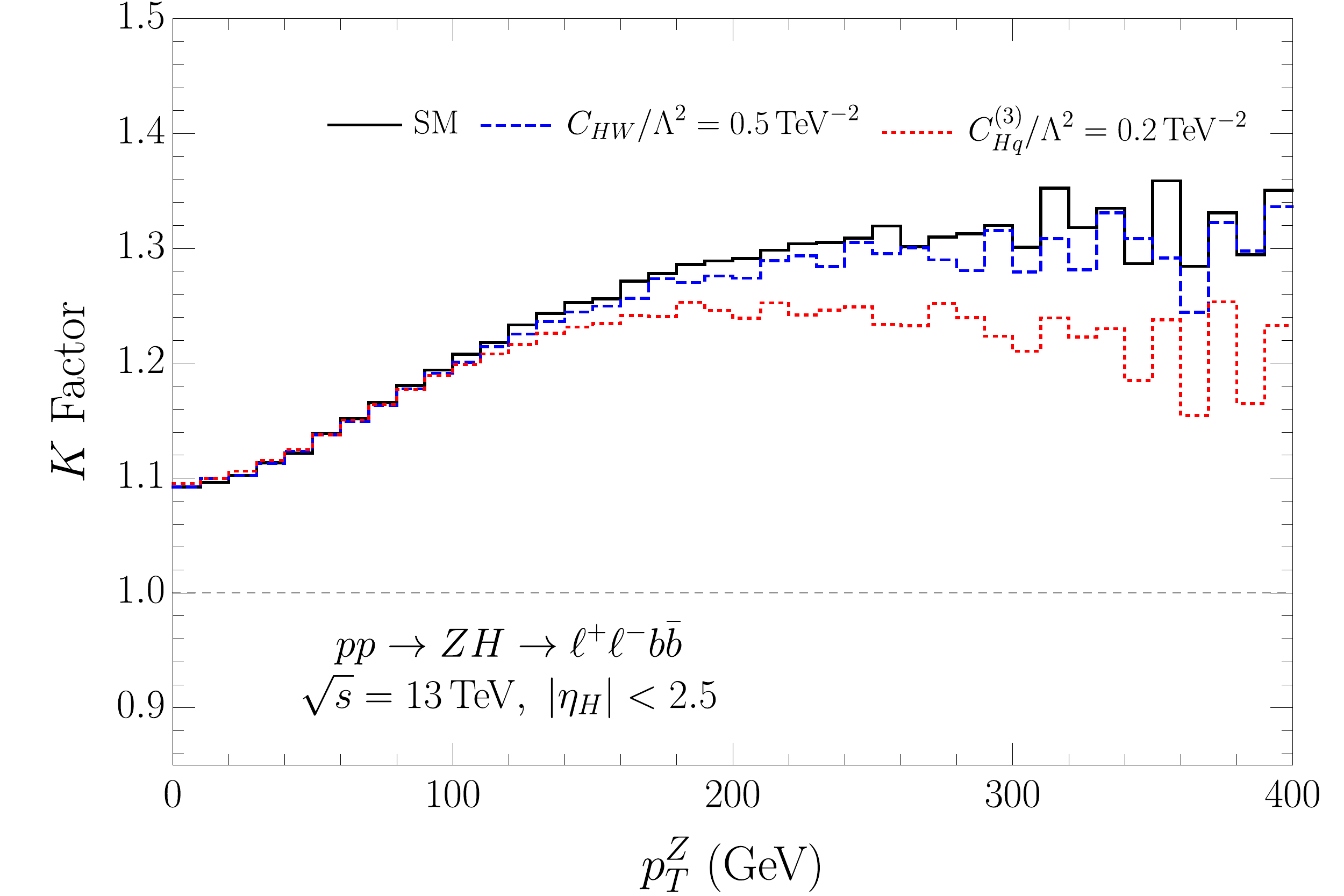}
  \caption{
Top: Differential distributions for $W^{\pm}H$ (left) and $ZH$ (right)
production at LO (dashed) and NLO (solid) in bins of $p_{T}^{V}$. We
plot the results for the SM (black), $C_{HW}/\Lambda^2 =
0.5~\mathrm{TeV}^{-2}$ (blue), and $C_{Hq}^{(3)}/\Lambda^2 =
0.2~\mathrm{TeV}^{-2}$ (red). 
Bottom: The associated $K$-factors for the same distributions at the
same three points, also in bins of $p_{T}^{V}$.  The figures are
computed to ${\cal{O}}(1/\Lambda^4)$.
}\label{fig:wh_ptw_dist}
\end{figure}

The effects of NLO QCD corrections in the SMEFT on distributions with
anomalous couplings in $W^+W^-$ and $W^\pm Z$ production have been
studied in previous work~\cite{Baglio:2017bfe, Baglio:2018bkm,
  Baglio:2019uty}. 
We now extend that analysis to include $W^\pm H$ and $ZH$
production~\cite{Alioli:2018ljm}.
For $W^+W^-$ production, it was demonstrated in Refs.~\cite{Baglio:2019uty,Baglio:2017bfe, Baglio:2018bkm} that the $K$-factor --- 
defined as the ratio of the NLO QCD (differential) cross section to
the LO one --- was largely unchanged by the presence of anomalous
gauge and fermion couplings.
For $W^\pm Z$ production, however, the effects of anomalous couplings
on the $K$-factor were found to be quite large. This can be understood
as the result of a delicate cancellation between the tree-level
diagrams leading to $W^\pm Z$ production in the SM, which are
intimately related to the presence of an approximate radiation
zero~\cite{Baur:1994ia} in the tree-level amplitude. 
The radiation zero is spoiled by the presence of QCD radiation,
leading to large $K$-factors~\cite{Baglio:2013toa} in some
differential distributions.
Because anomalous couplings affect the cancellation between the
tree-level diagrams, the interplay of radiation and anomalous
couplings makes an understanding of the NLO predictions crucial to
obtain accurate predictions of the distributions at the LHC.

We now consider the interplay of QCD corrections and anomalous
couplings on differential distributions for the associated production
of a Higgs and a $W^\pm$ or $Z$ gauge
boson~\cite{Campanario:2014lza,Granata:2017iod}.
While $W^\pm H$ and $ZH$ production do not have a tree-level radiation
zero as in $W^\pm Z$ production, the longitudinal modes at high energy
are closely related to the $W^+W^-$ and $W^\pm Z$ processes in the
high-energy limit by the Goldstone
theorem~\cite{Grojean:2018dqj,Panico:2017frx} and so we expect
interesting effects from QCD radiation.

In Fig.~\ref{fig:wh_ptw_dist}, we show the differential cross sections
for $W^\pm H$ and $ZH$ production in bins of $p_{T}^{V}$ for the
Standard Model and with $C_{HW}/\Lambda^2 = 0.5~\mathrm{TeV}^{-2}$
(blue) and $C_{Hq}^{(3)}/\Lambda^2 = 0.2~\mathrm{TeV}^{-2}$ (red).
This figure includes the differential cross sections evaluated to
${\cal{O}}({1\over\Lambda^4})$.
The NLO and LO predictions are shown as solid and dashed lines,
respectively. In the lower panels, we show the corresponding
$K$-factors at these benchmark points.
At both LO and NLO, we see that the effects of $C_{Hq}^{(3)}$ grow
fastest at high energy, due to the four-point interaction being
unsuppressed by an $s$-channel vector boson
propagator~\cite{Biekotter:2018rhp,Brehmer:2019gmn}. For the
anomalous-coupling points and for the SM, for both $W^\pm H$ and $ZH$
production, the $K$-factor becomes  larger at high $p_{T}^{V}$,
reaching $\sim 1.3$ for the SM at $400\,\mathrm{GeV}$.
While less pronounced than the effects in $W^\pm Z$ production,
treating the SMEFT contributions consistently at NLO  QCD 
in the SMEFT  changes the ratio of the NLO to LO predictions, and this
has an effect on the fits to the distributions as we show in
Section~\ref{sec:res}.

\subsection{Angular Distributions and Gauge Boson Polarizations at NLO} 

We now turn to a  discussion of the angular variables,
$\cos\theta_{W}^*$ of the decayed charged leptons in the gauge boson
rest frame. For $W^\pm Z$ production, we make use of the helicity
coordinate system defined by ATLAS in Ref.~\cite{Aaboud:2019gxl},
defining the $z$-direction of the $W^\pm$ rest frame by the $W^\pm$
direction in the diboson center-of-mass frame. More details are given
in Refs.~\cite{Bern:2011ie, Baglio:2018rcu}. For $W^\pm H$ and $ZH$
production, we use the same variables, with the $W^\pm H$ or $ZH$
system replacing the $W^\pm Z$ center-of-mass frame, and the
positively-charged lepton from the $Z$ decay playing the role of the
charged lepton in the $W$ frame.

These angular variables are useful because their distributions are
sensitive to the gauge boson polarizations~\cite{Panico:2017frx}.
They are of particular interest to us here because of the relationship
between the longitudinally polarized vector bosons and the Higgs boson.
Understanding the polarization fractions for high-energy vector bosons
has been shown to be a useful probe for anomalous-coupling
measurements at the LHC~\cite{Falkowski:2016cxu, Panico:2017frx,
  Franceschini:2017xkh}.
However, in Refs.~\cite{Campanario:2014lza,Baglio:2019uty} it was
found that the sensitivity of $\cos\theta_W^*$ to the anomalous
couplings was lost in the presence of an extra jet. This is a
manifestation of QCD corrections breaking the non-interference
between helicity amplitudes of the SM and the dimension-6 SMEFT
amplitudes, as originally pointed out in Ref.~\cite{Dixon:1993xd}, and
studied in the context of electroweak interactions in
Ref.~\cite{Azatov:2017kzw}.
Here, we consider the impact of vetoing
hard jets on restoring the sensitivity of these distributions at NLO.

\begin{figure}
  \includegraphics[width=0.49\linewidth]{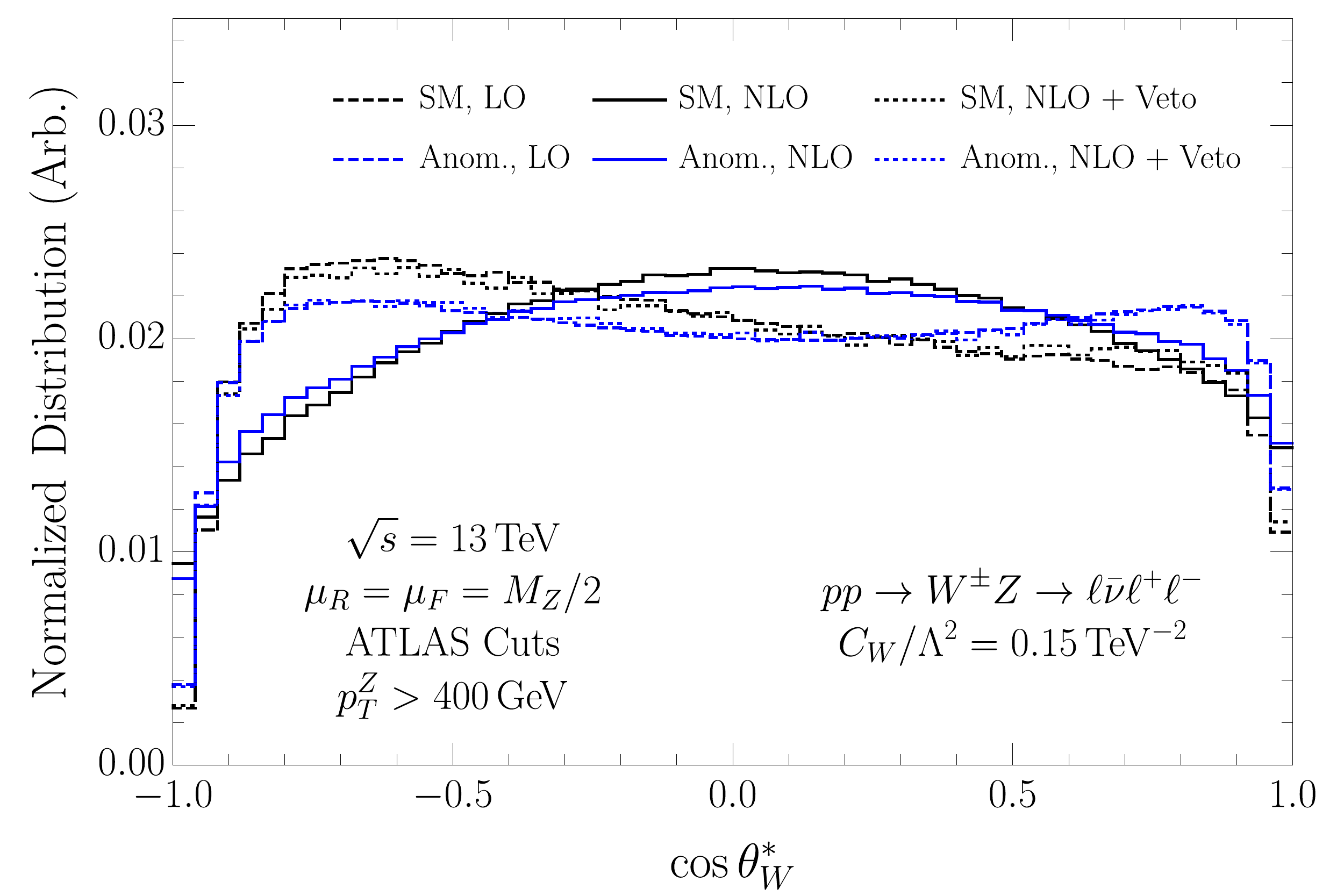}
  \hfill
  \includegraphics[width=0.49\linewidth]{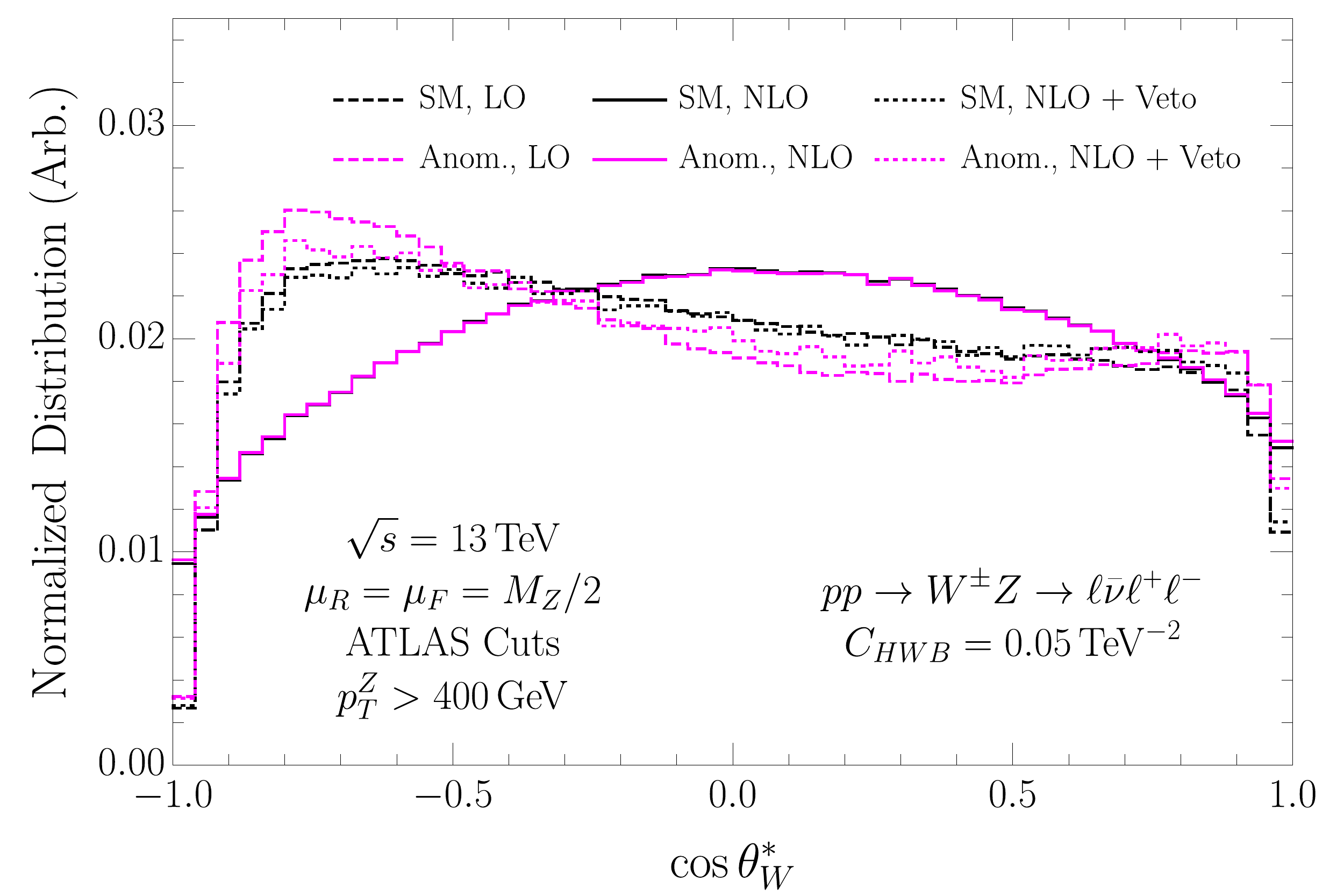}
  \vskip 0.25cm
  \includegraphics[width=0.49\linewidth]{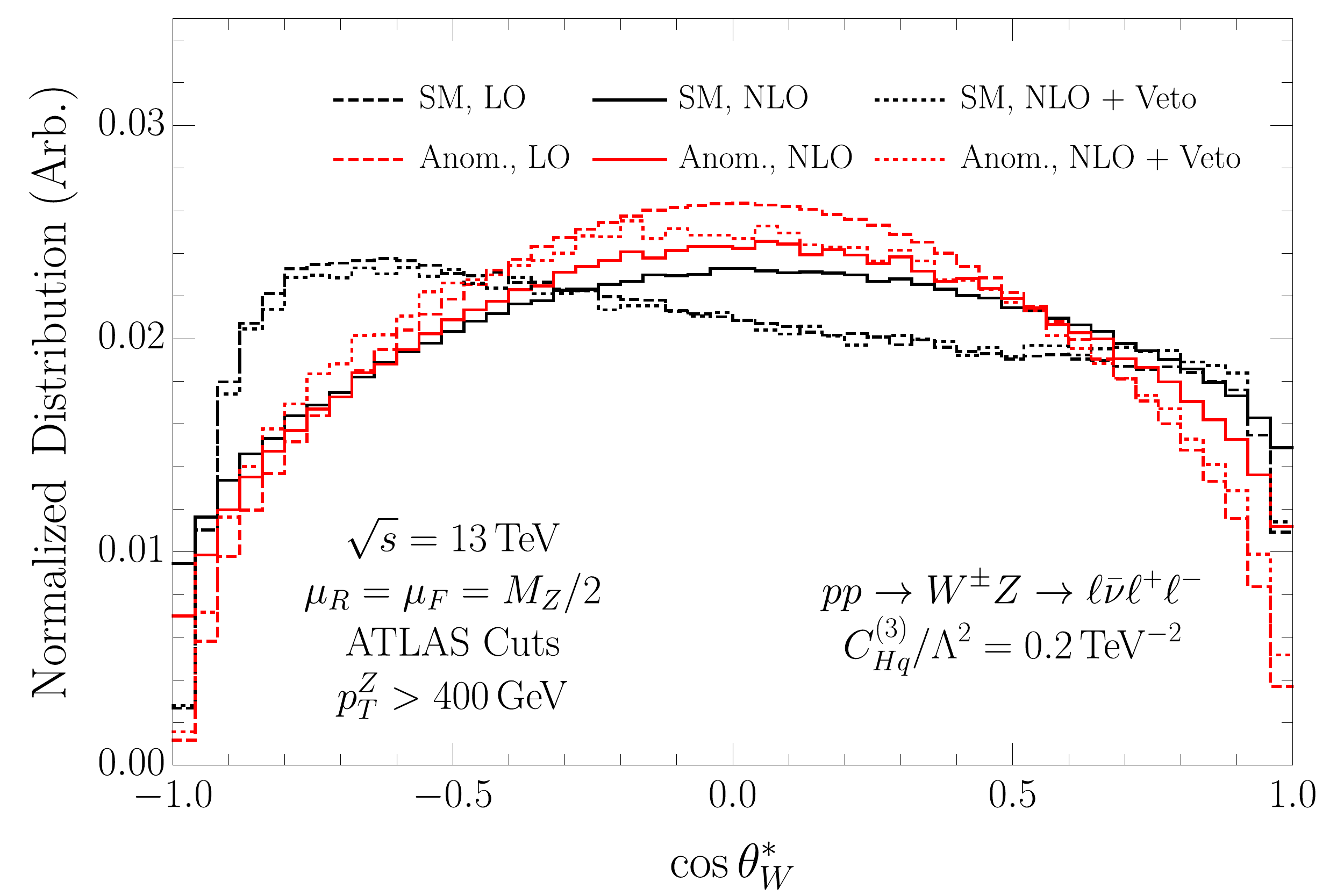}
  \vskip -0.25cm
  \caption{
    Normalized distributions of the angular variable, $\cos\theta_W^*$ in
    $WZ$ production at $13~\mathrm{TeV}$, requiring $p_{T}^{Z} >
    400\,\mathrm{GeV}$. In each figure we show the SM curve (black) for
    comparison along with an anomalous-coupling point: $C_W/\Lambda^2 =
    0.15~\mathrm{TeV}^{-2}$ (upper left), $C_{HWB}/\Lambda^2 =
    0.05~\mathrm{TeV}^{-2}$ (upper right), and $C_{Hq}^{(3)}/\Lambda^2 =
    0.2~\mathrm{TeV}^{-2}$ (bottom). The points are chosen to be near the
    edge of the allowed regions by our combined fits to $WV$ and $VH$
    production ($V=W^\pm,Z$). For each parameter point, we plot the
    distribution at LO (dashed), NLO (solid), and at NLO with a jet veto
    (dotted). The jet veto curves correspond to a veto on jets with
    $p_{T}^{j} > 150\,\mathrm{GeV}$. The figures are 
    computed to ${\cal{O}}(1/\Lambda^4)$.
  }\label{fig:wz_costhetaw}
\end{figure}

In Fig.~\ref{fig:wz_costhetaw}, we present the normalized
$\cos\theta_W^*$ distributions from $W^{\pm}Z$ production at LO, NLO
and at NLO with a $150\,\mathrm{GeV}$ jet veto. In all plots we also
include a $p_T^Z$ cut, $p_T^Z>400$~GeV, in order to enhance our
sensitivity to the anomalous couplings. In each figure we show the
results for the SM as well as with one of three anomalous couplings:
$C_W/\Lambda^2 = 0.15~\mathrm{TeV}^{-2}$ (upper left),
$C_{HWB}/\Lambda^2 = 0.05~\mathrm{TeV}^{-2}$ (upper right), and
$C_{Hq}^{(3)}/\Lambda^2 = 0.2~\mathrm{TeV}^{-2}$ (bottom).
As is clear from comparing the LO (dashed) and NLO (solid) curves, the
hard radiation present  in $W^\pm Z$ production at NLO washes out much
of the sensitivity to anomalous couplings, as the SM and
anomalous-coupling curves are essentially indistinguishable at NLO,
despite the differences at LO.
With a veto on hard jets, however, the sensitivity is restored,
essentially to the levels obtained at LO. At high energy, only the
$(00)$ polarization (where both gauge bosons are longitudinally
polarized) and $(\pm,\mp)$ (transverse polarizations)
survive~\cite{Baur:1994ia}, and the angular distributions of the
polarizations are different.
The longitudinal polarization amplitude receives no contribution from
the anomalous gauge couplings in the high-energy limit. 
Furthermore, only  $C_W$ contributes to the high-energy limit of the
$(\pm,\mp)$ amplitude. 
We also note that when only $C_{HWB}$ is turned on (pink curves, upper
right in Fig.~\ref{fig:wz_costhetaw}), the anomalous couplings
$\lambda^Z = \lambda^{\gamma}$ are fixed to zero. Here, we can see
that  with the smaller value of $C_{HWB}$, the transverse contribution
(which peaks at large $|\cos\theta_W^*|$) is enhanced, and the process
is more sensitive to the $\delta g_1^Z$, $\delta\kappa^{\gamma}$
deviations than to the $\lambda^Z = \lambda^{\gamma}$ anomalous
couplings.

\begin{figure}
  \includegraphics[width=0.49\linewidth]{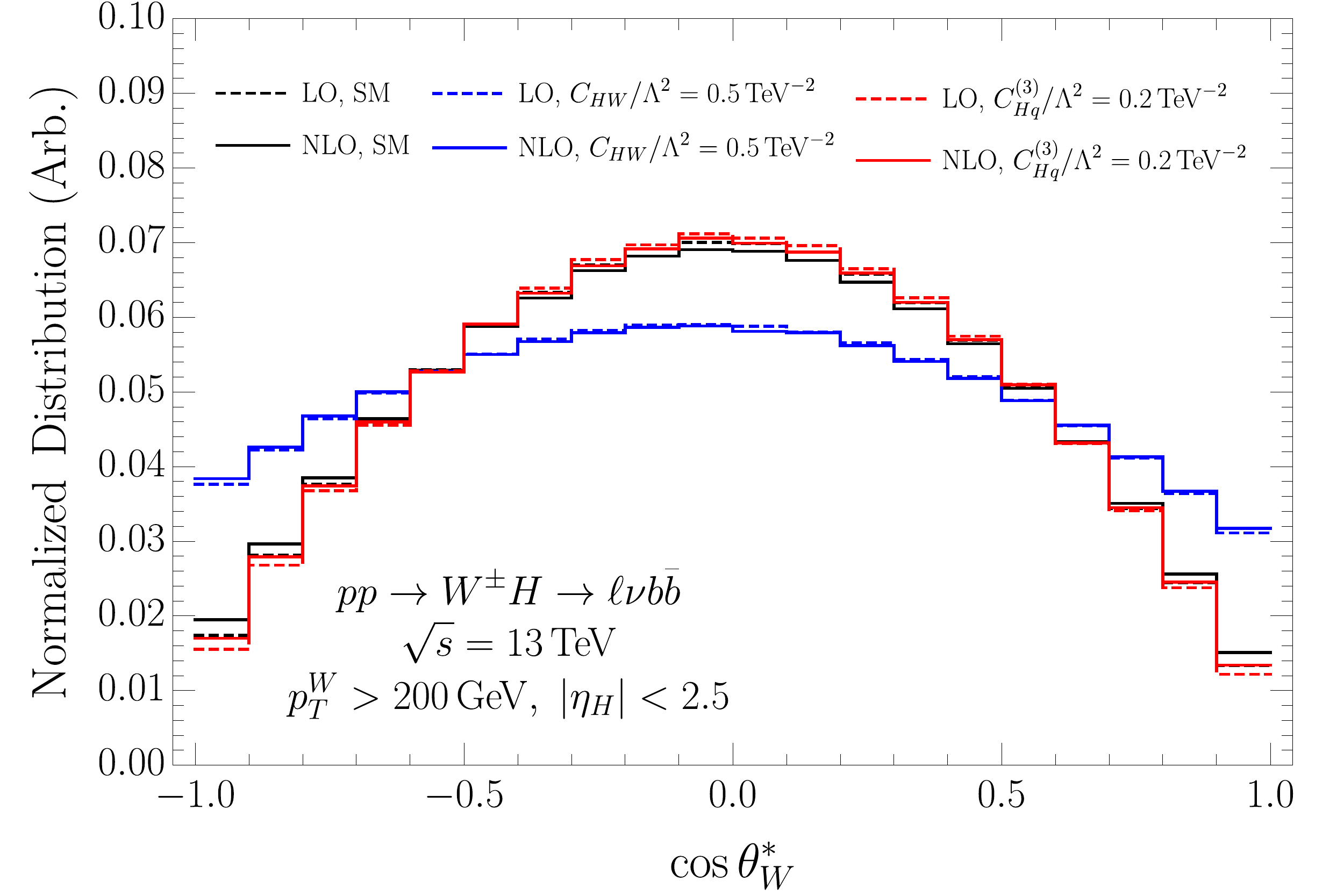}
  \hfill
  \includegraphics[width=0.49\linewidth]{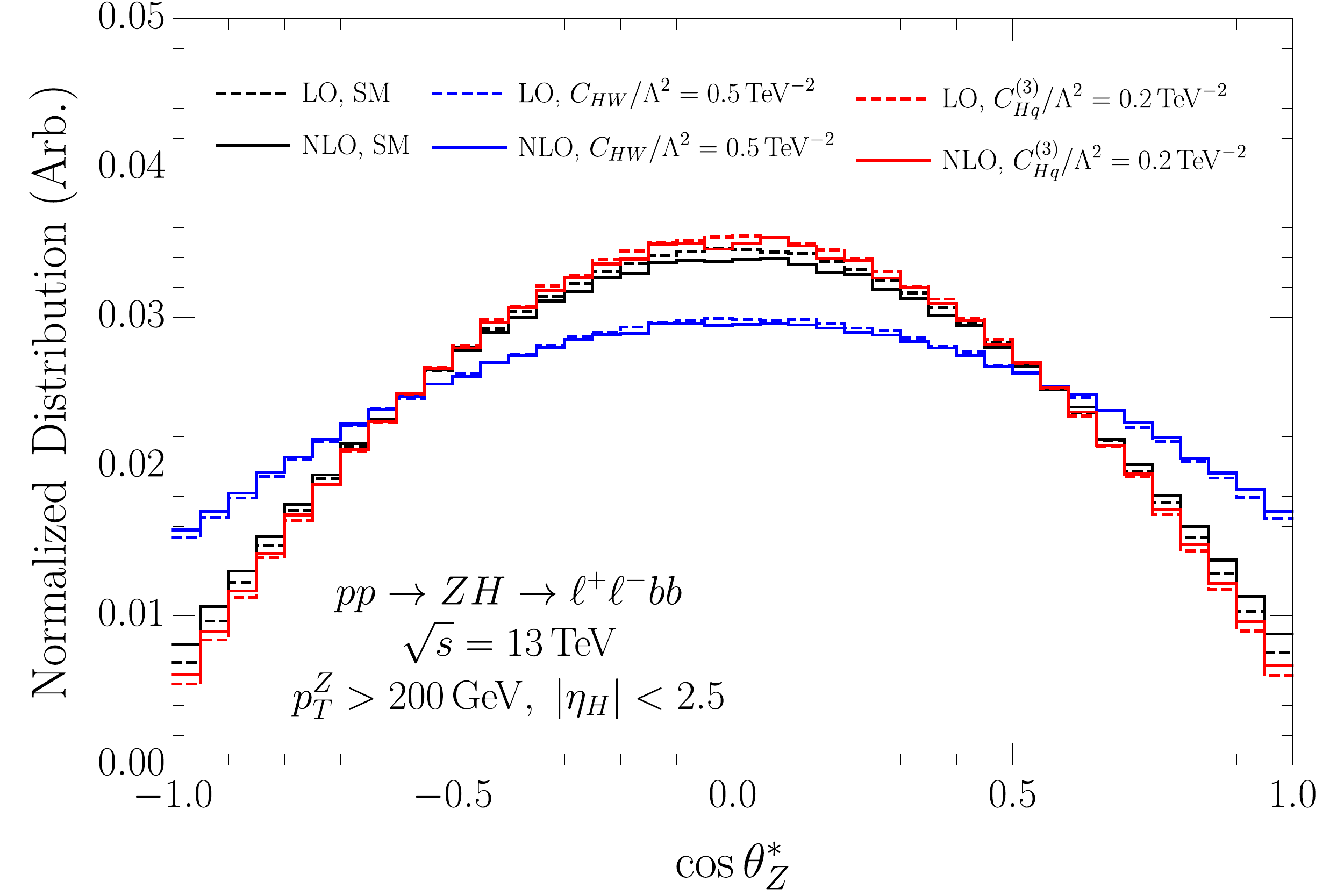}
  \caption{
    Normalized distributions of the angular variable $\cos\theta_W^*$
    in $WH$ production (left) and $\cos\theta_Z^*$ in $ZH$ production
    (right), both at $13~\mathrm{TeV}$ and requiring the vector boson
    to have $p_{T}^{V} > 200\,\mathrm{GeV}$ and the Higgs rapidity to
    lie within $\pm 2.5$.
    In each figure, we present the results at LO (dashed) and NLO
    (solid) for three different parameter points: the SM (black), with
    $C_{HW}/\Lambda^2 = 0.5~\mathrm{TeV}^{-2}$ (blue), and with
    $C_{Hq}^{(3)}/\Lambda^2 = 0.2~\mathrm{TeV}^{-2}$ (red). The
    figures are computed to ${\cal{O}}(1/\Lambda^4)$.
  }\label{fig:wh_costhetaw}
\end{figure}

In Fig.~\ref{fig:wh_costhetaw}, we show the normalized distributions
of the analogous angular variables but for $W^\pm H$ and $ZH$
production. We plot the results at LO (dashed) and NLO (solid) for
$C_{HW}/\Lambda^2 = 0.5~\mathrm{TeV}^{-2}$ (blue) and
$C_{Hq}^{(3)}/\Lambda^2 = 0.2~\mathrm{TeV}^{-2}$ (red). Here, we see
that with $C_{Hq}^{(3)}$ nonzero, the distribution has a very similar
shape to the SM piece, as both are dominated by the longitudinally
polarized helicity amplitudes at high $p_T$.
The distribution with nonzero $C_{HW}$, however, enhances the
transverse parts of the amplitude, and thus has a shape that is
enhanced at $\cos\theta_V^* = \pm 1$. This can be clearly seen in the
LO results of Ref.~\cite{Brehmer:2019gmn}. In contrast to $W^\pm Z$,
these distributions are largely unchanged by the higher-order
corrections, and maintain their sensitivity to anomalous couplings
that enhance the transverse polarizations even in the presence of
radiation.

\subsection{Sensitivity to Anomalous Couplings}

\begin{figure}[htp!]
\includegraphics[width=0.49\linewidth]{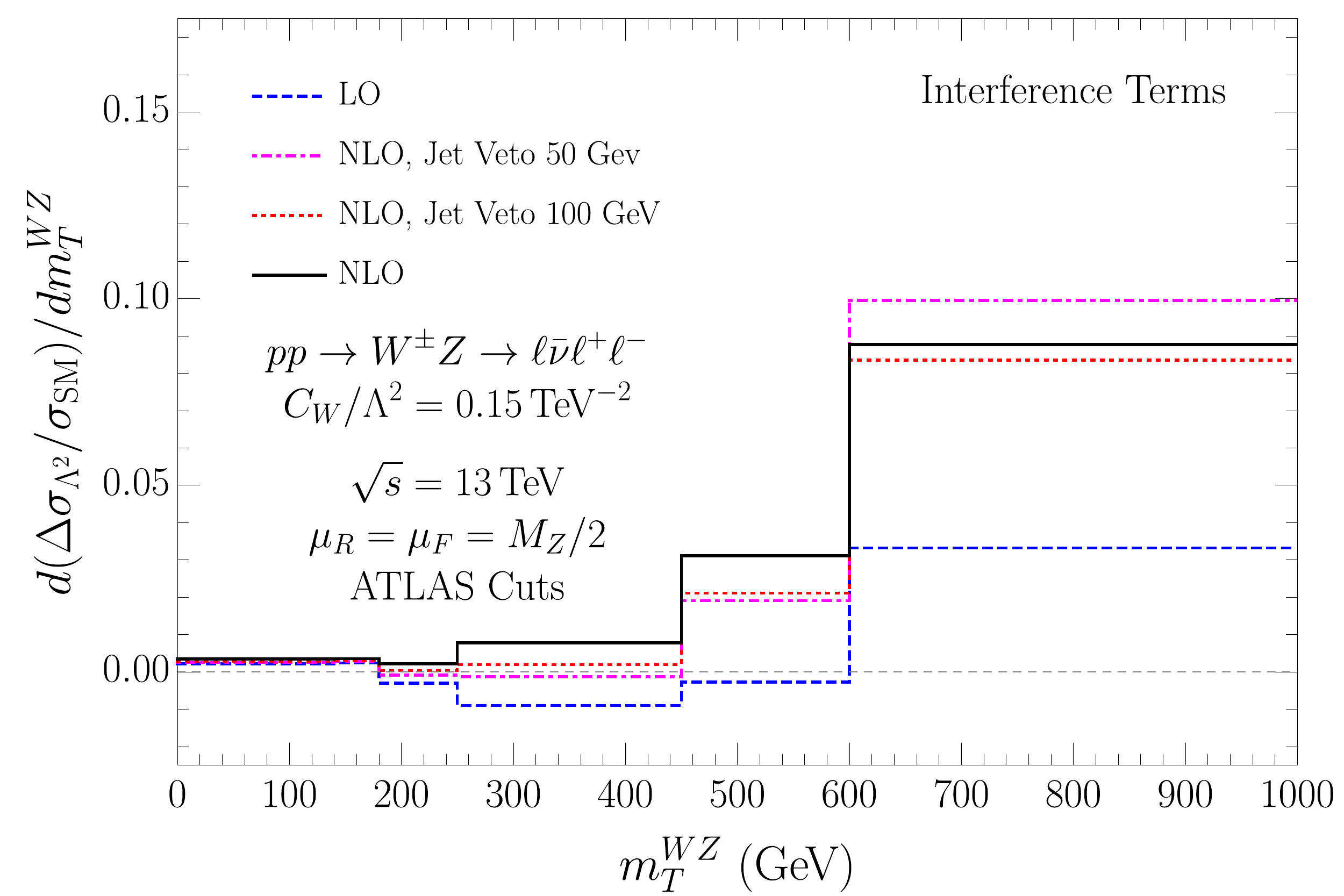}
\hfill %
\includegraphics[width=0.49\linewidth]{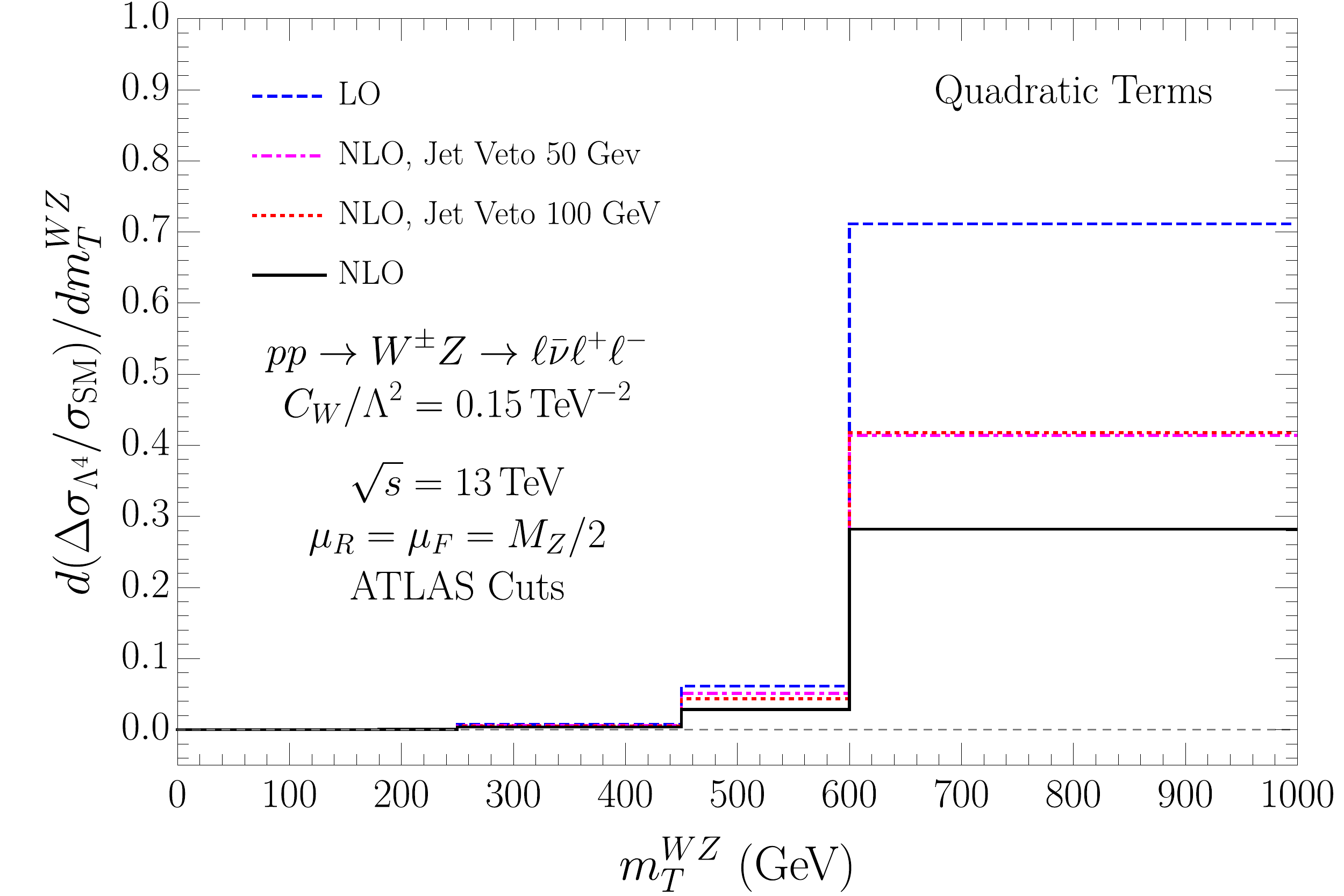}\\
\vskip 0.2cm
\includegraphics[width=0.49\linewidth]{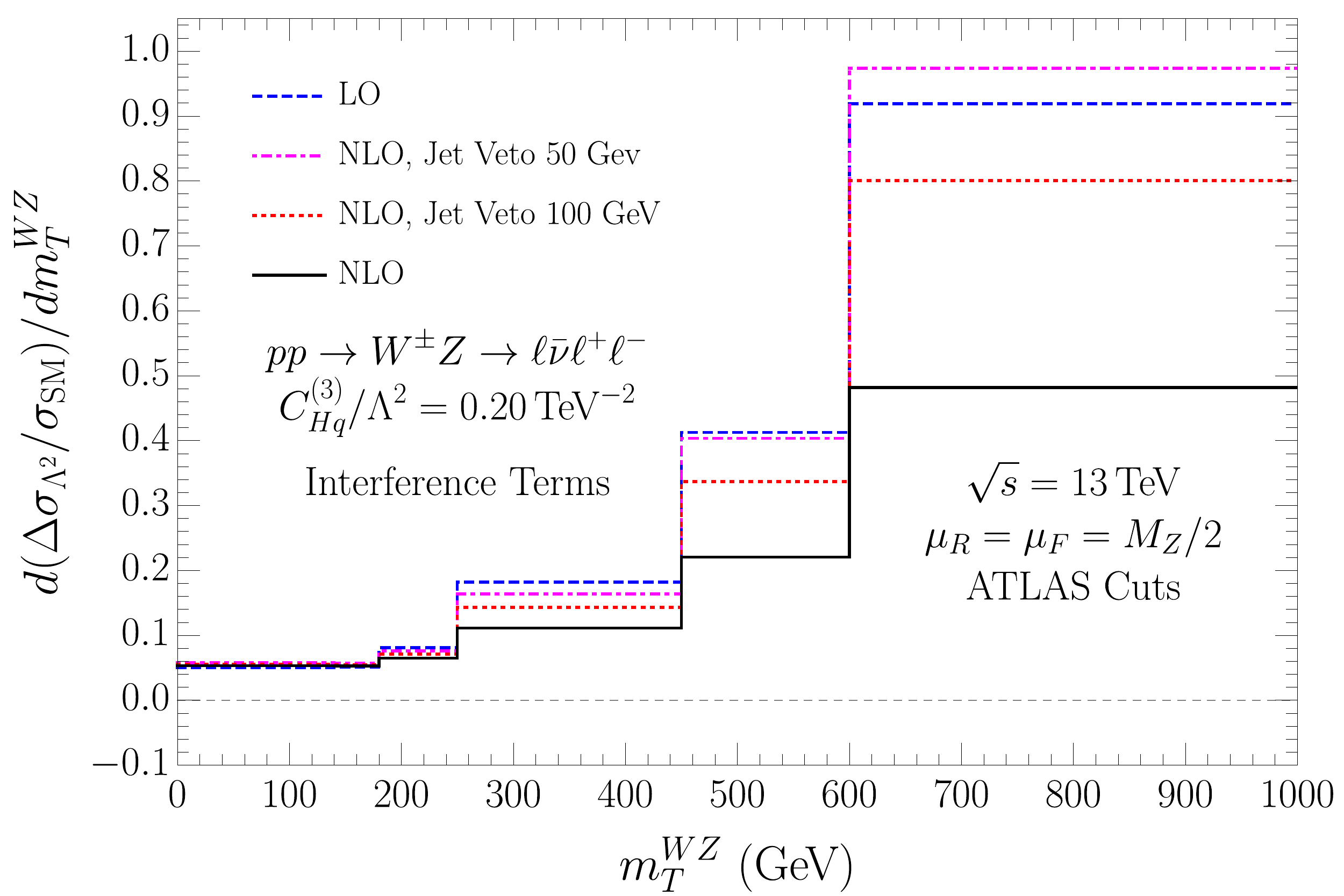}
\hfill %
\includegraphics[width=0.49\linewidth]{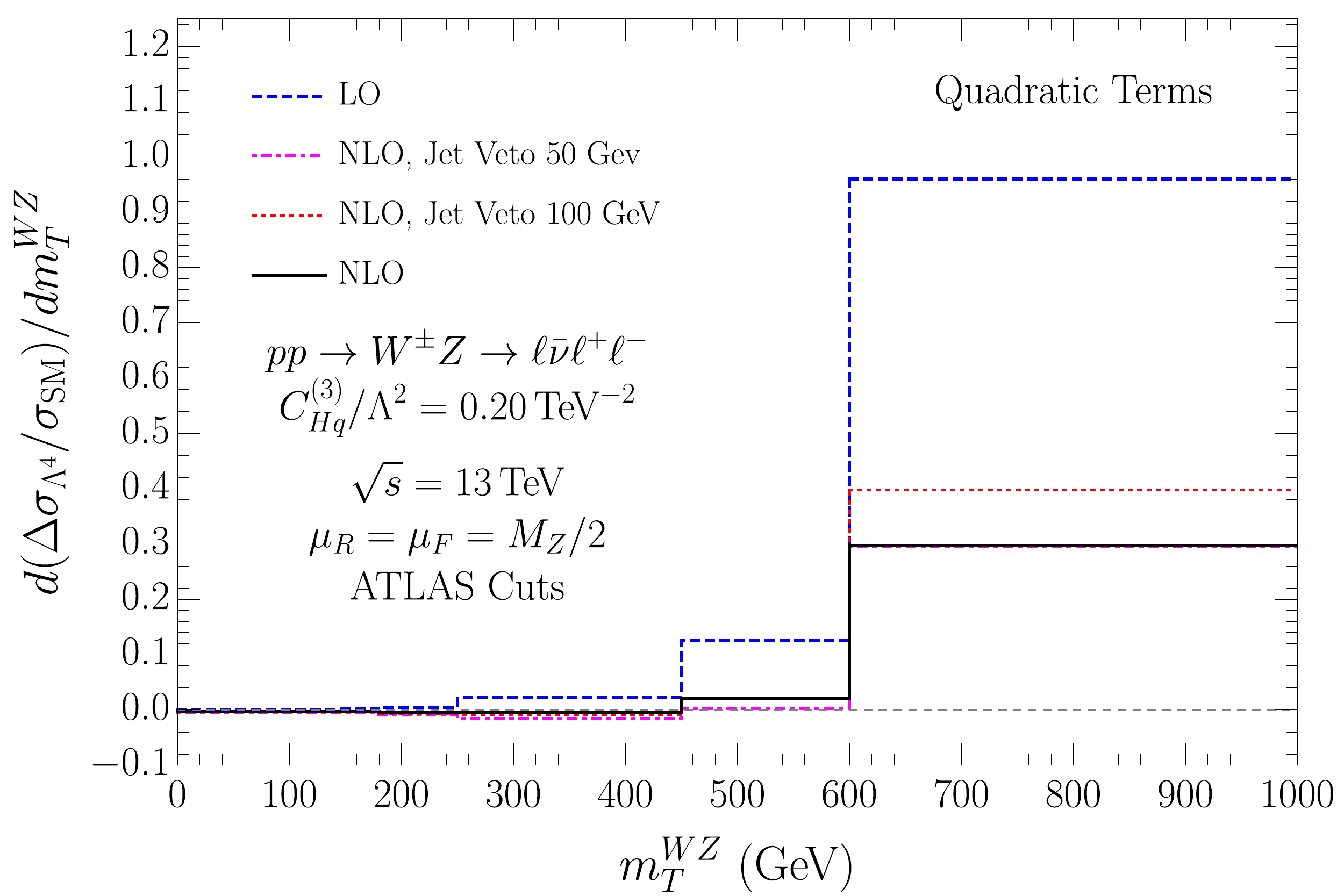}
\caption{
The sensitivity of the $m_{T}^{WZ}$ distributions to anomalous gauge
(top) and fermion (botton) couplings, at LO and NLO with varying jet
vetos. See text for details.
}\label{fig:deltasig_wz}
\end{figure}

We can also consider how the jet veto changes the sensitivity to
anomalous couplings in other  $W^\pm Z$ distributions. If we decompose
a generic differential cross section up to $\mathcal{O}(\Lambda^{-4})$
as
\begin{equation}
\sigma(C_i) = \sigma_{\textrm{SM}} + \Delta\sigma_{\Lambda^2}(C_i) + \Delta\sigma_{\Lambda^4}(C_i^2),
\label{eq:dec}
\end{equation}
we can isolate parts of the cross section that depend linearly and
quadratically on the Wilson coefficients, and see how these parts grow
with energy at LO, and in the presence of radiation. This is done in
Fig.~\ref{fig:deltasig_wz} for $W^\pm Z$ production with $C_{W} /
\Lambda^2 = 0.15~\mathrm{TeV}^{-2}$ (top) and $C_{Hq}^{(3)} /\Lambda^2
= 0.2~\mathrm{TeV}^{-2}$ (bottom) in bins of $m_{T,WZ}$ for the
$\Lambda^{-2}$ (left) and $\Lambda^{-4}$ (right) pieces,
respectively.

We see immediately that the presence of QCD radiation makes a
substantial difference in the sensitivity of the distributions to
anomalous couplings. 
Focusing first on the linear pieces, we note that these arise from the
interference between the dimension-6 SMEFT part of the amplitude with
the SM part, and are thus subject to the non-interference effects
noted in Refs.~\cite{Falkowski:2016cxu, Panico:2017frx,
  Franceschini:2017xkh}.
At high energies, the SM amplitude receives contributions from both
longitudinally and oppositely-polarized transverse gauge bosons. The
portion of the amplitude proportional to $C_W$, however, has only
transverse polarizations.
The resulting non-interference between the SM and the dimension-6
SMEFT  amplitudes is clear from the blue curve in
Fig.~\ref{fig:deltasig_wz} (upper left), which  does not substantially
grow with energy.
As discussed in Ref.~\cite{Azatov:2017kzw}\footnote{This was
  originally pointed out in a slightly different context in
  Ref.~\cite{Dixon:1993xd}.}, however, the presence of an extra 
quark or gluon in the matrix element allows for this interference to
be restored, and indeed, we see that the interference term at NLO
(with or without a jet veto) grows substantially at high $m_{T}^{WZ}$.
That this enhanced sensitivity to the interference persists even with
a veto on the hard jets arising from the real emission implies that
the virtual corrections play an important role in restoring the
interference.

For the interference term proportional to $C_{Hq}^{(3)}$, the story is
somewhat different.
Here, we see that there is a growth in sensitivity at high energies
even at LO, as the $C_{HQ}^{(3)}$ amplitude enhances the longitudinal
parts of the amplitude which are already dominant in the SM part at
high energies.
At NLO, much of this sensitivity is washed out due to the presence of
hard jets, but a great deal of the sensitivity can be restored by
imposing a veto on the hard real emission.

Turning now to the $\mathcal{O}(\Lambda^{-4})$ terms, we see
immediately on the right hand side of Fig. ~\ref{fig:deltasig_wz} that
the LO distributions exhibit much faster growth with energy than the
corresponding NLO curves, both for $C_W$ and $C_{Hq}^{(3)}$.
The $\mathcal{O}(\Lambda^{-4})$ terms do not depend on any
interference with the SM amplitude, so the sensitivity is dictated
largely by the kinematics of the process.
For anomalous gauge couplings, this was studied in
Ref.~\cite{Campanario:2014lza}, where it was found that $W^\pm Z$
production at NLO generically allows for hard jets, which suppresses
the sensitivity to the anomalous-coupling pieces (which grow like
$({\text{Energy}})^2$).
It was found there that much of the sensitivity in this distribution
can be regained by vetoing events containing hard jets.
The same conclusion is apparent both for $C_W$ and $C_{Hq}^{(3)}$ in
Fig.~\ref{fig:wz_costhetaw}, where vetoing jets with $p_{T} >
150\,\mathrm{GeV}$ restores much of the sensitivity obtained at LO.

In principle, one could perform the same analysis on the
$\mathcal{O}(\Lambda^{-2})$ and $\mathcal{O}(\Lambda^{-4})$ terms in
the $W^\pm H$ distributions.
In practice, though, the results are significantly less interesting when comparing LO to NLO.
This is because, as shown in Ref.~\cite{Campanario:2014lza}, the real
emission contributions to $W^\pm H$ production at NLO are typically
soft, in contrast to the hard jets that appear in $W^\pm Z$
production.
Thus, a veto on hard jets  in $W^\pm H$ production at NLO does not
significantly change the sensitivity to anomalous couplings, either at
$\mathcal{O}(\Lambda^{-2})$ or
$\mathcal{O}(\Lambda^{-4})$. Furthermore, as can be seen in
Fig.~\ref{fig:wh_ptw_dist}, the $K$-factors are only mildly dependent
on the anomalous couplings, so the sensitivity at LO and NLO to all
higher-dimension operators is largely the same.

\section{Fits to Warsaw Coefficients}
\label{sec:res}

\subsection{Datasets and Fitting Procedure}

Section~\ref{sec:dists}  demonstrates that the implementation of NLO
QCD within the SMEFT can have  a significant impact on
distributions. These changes lead to different predictions from those
obtained by using LO QCD in the SMEFT with the appropriate Standard
Model $K$-factor. We further solidify the need to include NLO QCD
within SMEFT fits by showing the differences between fits with and
without NLO. We also show that including $\mathcal{O}(1/\Lambda^4)$
can significantly improve the fits. Lastly, the
$\mathcal{O}(1/\Lambda^4)$ terms allow one to explore if the values of
coefficients are consistent with a weakly- or strongly-coupled theory.

We fit to the $10$ Warsaw basis coefficients described in
Sect. \ref{sec:basics}  at both LO and NLO in the SMEFT to quantify
these effects. We calculate uncorrelated $\chi^2$ fits to differential
cross section measurements for the processes $W^\pm H, ZH, W^+W^-$,
and $W^\pm Z$ and we construct the $\chi^2$  function for a given
anomalous-coupling input, $\vec{C}$, as
\begin{align}
\chi^2(\vec{C}) = 
\sum_{\substack{\text{\it WH, ZH} \\ \text{\it WW, WZ}} }^{\text{processes}} 
\sum_{\alpha}^{\text{datasets}} 
\sum_{i}^{\text{bins}} 
\frac{(\epsilon_{i \alpha} \hat{O}(\vec{C})_{i \alpha}^{\text{theory}} - \hat{O}_{i\alpha}^{\text{exp}})^2 }{
(v^{\text{exp}}_{i \alpha})^2}\, ,
\end{align}
where  $\hat{O}(\vec{C})_{i\alpha}^{\text{theory}}$,
$\hat{O}_{i\alpha}^{\text{exp}}$, and $v^{\text{exp}}_{i\alpha}$ are
respectively the theoretical expected value,  experimental
observation, and estimated uncertainties for the $i^{\text{th}}$ bin
of dataset $\alpha$.  An  efficiency factor, $\epsilon_{i\alpha}$, is
introduced to account for an overall scaling of the simulation data,
where $\epsilon_{i\alpha}$ is calculated by taking the ratio of the
experimentally simulated value for the SM differential cross section
over our prediction for the differential cross section with an SM
input ($\vec{C} = 0$) for the $i^{\text{th}}$ bin of dataset $\alpha$.

The datasets that go into each process are detailed in
Table~\ref{tab:data}. The uncertainties are estimated by combining
reported statistical and systemic uncertainties in quadrature,
assuming an overall 5\% systematic uncertainty bin-by-bin, neglecting
correlations.
\newcommand{\sla}[1]{/\!\!\!\!#1}
\begin{table}
\begin{center}
\begin{tabular}{ l|lcll}
\hline
Channel  & Distribution & \# bins   &\hspace*{0.2cm} Data set & \hspace*{0.2cm}Int.  Lum.
\\[0mm]
\hline
$W^\pm H\rightarrow b \bar{b} \ell^{\pm}+\sla{E}_T$ & $p^{W}_{T}$, Fig.~3 & 2 
& ATLAS 8~TeV &79.8 fb$^{-1}$~\cite{Aaboud:2019nan}
\\[0mm]
$ZH\rightarrow b \bar{b}\ell^+\ell^{-} \text{ or }b \bar{b} +\sla{E}_T $ & $p^{Z}_{T}$, Fig.~3 & 3 
& ATLAS 8~TeV &79.8 fb$^{-1}$~\cite{Aaboud:2019nan}
\\[0mm]
\hline
$W^+ W^- \rightarrow \ell^+\ell^{\prime -}+\sla{E}_T\; (0j)$ & $p^{\rm leading, lepton}_{T}$, Fig.~11 & 1 
& ATLAS 8~TeV &20.3 fb$^{-1}$~\cite{Aad:2016wpd}
\\[0mm]
$W^+ W^- \rightarrow e^\pm \mu^\mp+\sla{E}_T\; (0j)$ &  $p_T^{\rm leading ,lepton}$, Fig.~7 & 5 
& ATLAS 13~TeV &36.1 fb$^{-1}$~\cite{Aaboud:2019nkz}
\\[0mm]
\hline
$W^\pm Z\rightarrow \ell^+\ell^{-}\ell^{(\prime)\pm}$ & $m_{T}^{WZ}$, Fig.~5 & 2 
& ATLAS 8~TeV & 20.3 fb$^{-1}$~\cite{Aad:2016ett}
\\[0mm]
$W^\pm Z\rightarrow \ell^+\ell^{-}\ell^{(\prime)\pm}+\sla{E}_T$ 
& $Z$ candidate $p_{T}^{\ell\ell}$, Fig.~5 & 9 
& CMS 8~TeV &19.6 fb$^{-1}$~\cite{Khachatryan:2016poo}
\\[0mm]
$W^\pm Z\rightarrow \ell^+\ell^{-}\ell^{(\prime)\pm}$ &  $m_{T}^{WZ}$ Fig.~4c & 6
& ATLAS 13~TeV &36.1 fb$^{-1}$~\cite{Aaboud:2019gxl}
\\[0mm]
$W^\pm Z\rightarrow \ell^+\ell^{-}\ell^{(\prime)\pm}+\sla{E}_T$ & $m^{WZ}$, Fig.~15a & 3
& CMS 13~TeV, &35.9 fb$^{-1}$~\cite{Sirunyan:2019bez}
\\[0mm]
\hline
\end{tabular}
\caption{
Experimental data included in our study. The third column shows the
number of bins used in our analysis, always counting from the
highest.
}\label{tab:data}
\end{center}
\end{table}

We explore two methods for calculating confidence intervals of the
Warsaw coefficients: projecting all but one coefficient to zero and
alternatively profiling over the remaining coefficients to minimize
the $\chi^2$  function at each point. The numerical results obtained
by fitting all\footnote{The fits to individual processes can by
  compared in Tables~\ref{tab:WHZHfits},~\ref{tab:WWWZfits}, and
  \ref{tab:WVVHfits} located in the Appendix.} processes using both
profiling and projecting are given in Table~\ref{tab:fits}. They are
compared graphically in Figures~\ref{fig:1Dproject} and
\ref{fig:1Dprofile}. Overall we see that the projected limits are
significantly more stringent than the profiled. This is to be expected
since the profiling allows for more flexibility in the $\chi^2$
function. The profiling method demonstrates the multidimensional
nature of the fit.

We also show several 2D confidence interval fits using  the projection
method in Figure~\ref{fig:2Dplots}. In principle one could make a 2D
confidence interval for each combination of Warsaw
coefficients. However, most of these plots end up with similar
results, showing  order 20\% NLO effects  and with many of the regions
falling in the strongly-coupled regime. We have selected some example
plots that are particularly demonstrative and also correspond to
interesting electroweak precision variables (S and T).

\begin{table}
\begin{tabular}{|c|c|c|c|c|c|c|c|c|}
\hline
 \text{} & \multicolumn{4}{c|}{$W^{+}W^{-}+W^{\pm}Z + ZH + W^{\pm}H$ Projected}
 	   &   \multicolumn{4}{c|}{$W^{+}W^{-}+W^{\pm}Z + ZH + W^{\pm}H$ Profiled} \\
 \cline{2-9}
 & \multicolumn{2}{c|}{$\Lambda^{-4}$} & \multicolumn{2}{c|}{$\Lambda^{-2}$} & 
     \multicolumn{2}{c|}{$\Lambda^{-4}$} & \multicolumn{2}{c|}{$\Lambda^{-2}$} \\
 \cline{2-9}
  & \text{LO }& \text{NLO }& \text{LO } & \text{NLO } 
  & \text{LO }& \text{NLO}& \text{LO }& \text{NLO } \\
\hline
$ C_{\text{HWB}}$ 
& (-.05, .03) & (-.09, .04) & (-.07, .02) & (-.14, .03) 
& (-.70, .47) & (-.75, .50) & (-2.9, 2.3) & (-4.4, 1.5)
\\
$ C_{\text{Hq}}^{(3)}$ 
& (-.02, .08) & (-.02, .11) & (-.02, .09) & (-.02, .14) 
& (-.26, .62) & (-.30, .67) & (-.17, .82) & (-.38, .82)
\\
$ C_{\text{HD}} $
& (-.12, .06) & (-.21, .08) & (-.15, .05) & (-.30, .07) 
& (-1.1, 2.1) & (-1.2, 2.4) & (-4.5, 6.8) & (-2.6, 9.1)
\\
$ C_{\text{Hq}}^{(1)}$ 
& (-.16, .21) & (-.18, .19) & (-.24, .20) & (-.32, .15) 
& (-.21, .38) & (-.25, .40) & (-.45, .93) & (-.81, .71)
\\
$ C_{\text{Hu}} $
& (-.30, .22) & (-.33, .24) & (-.34, .72) & (-.38, .81) 
& (-.43, .59) & (-.46, .62) & (-23., 23.) & (-42., 48.)
\\
 $C_{\text{HW}}$ 
 & (-1.1, .55) & (-1.2, .56) & (-.52, .92) & (-.52, .92) 
 & (-1.4, 2.4) & (-1.5, .51) & (-31., 19.) & (-33., 17.)
\\
 $C_W $
 & (-.13, .13) & (-.20, .18) & (-1.4, 1.3) & (-.28, .93) 
 & (-.14, .14) & (-.20, .19) & (-1.3, 1.9) & (-3.2, 2.1) 
 \\
 $C_{\text{Hd}} $
 & (-.31, .35) & (-.33, .38) & (-2.2, 1.1) & (-2.2, 1.0) 
 & (-.62, .45) & (-.67, .48) & (-82., 86.) & (-13., 14.)
\\
 $C_{H\square } $
 & (-4.9, 6.3) & (-4.9, 6.3) & (-4.6, 8.6) & (-4.6, 8.6) 
 & (-57., 20.) & (-59., 20.) & (-27., 43.) & (-25., 43.)
\\
 $C_{\text{HB}}$ 
 & (-2.8, 2.3) & (-2.9, 2.4) & (-6.1, 11.) & (-6.0, 12.) 
 & (-3.1, 3.8) & (-3.3, 4.0) & (-31., 22.) & (-31., 21.)
\\
\hline
\end{tabular}
\caption{$95 \%$ confidence interval fits to individual EFT
  coefficients using $W^{+}W^{-}+W^{\pm}Z + ZH + W^{\pm}H$ data, with
  $\Lambda$ fixed to $1~\mathrm{TeV}$.}
\label{tab:fits}
\end{table}

\subsection{Importance of NLO QCD and Quadratic Order Fits}

\begin{figure}
  \centering
  \includegraphics[width=.8\textwidth]{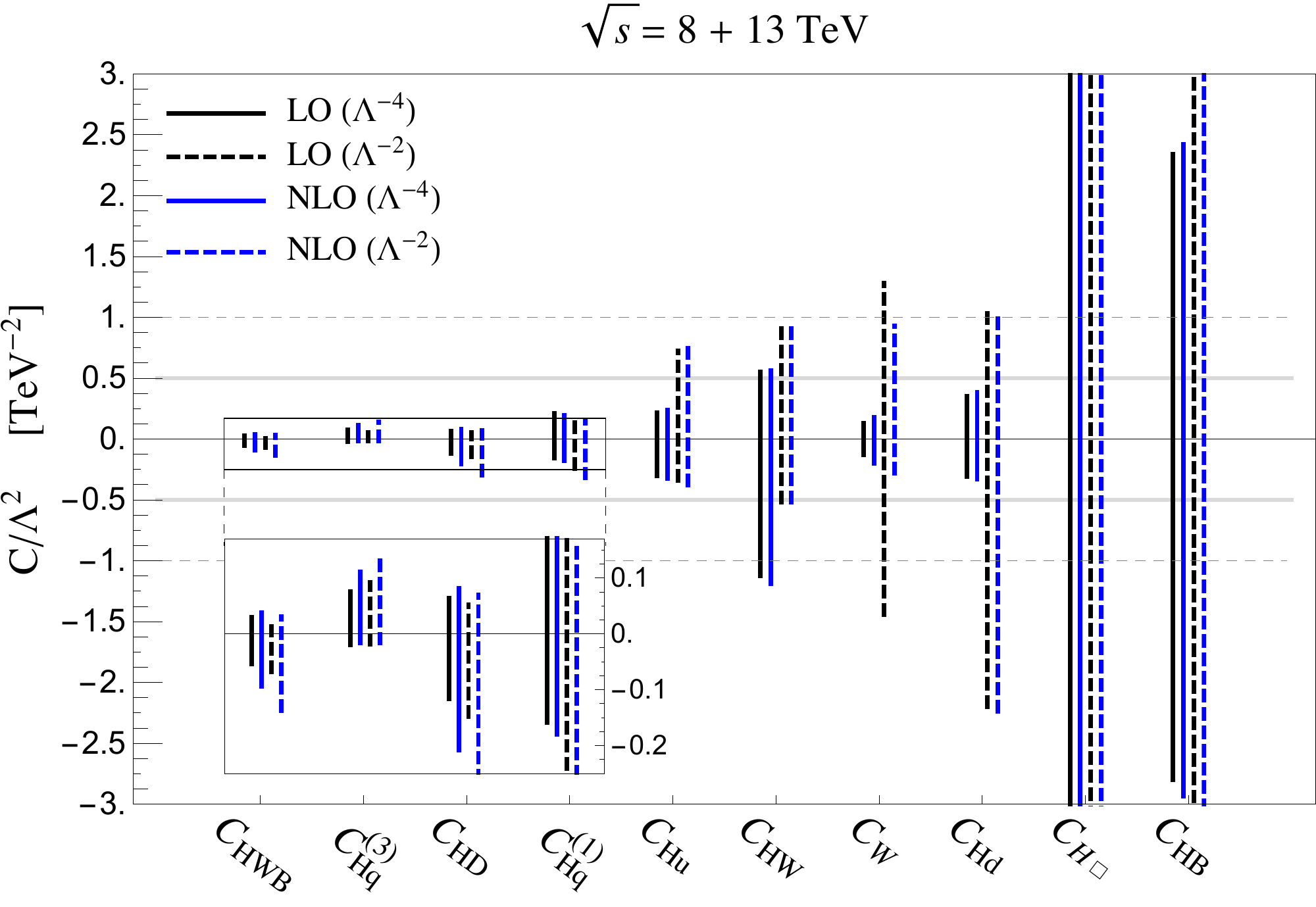}
  \caption{$95 \%$ confidence interval fits to individual EFT
    coefficients using $W^{+}W^{-}+W^{\pm}Z + ZH + W^{\pm}H$
    data. All other SMEFT coefficients are projected to
    zero. Fits quadratic  in $1/\Lambda^2$ (Linear) LO and NLO
    QCD are shown as solid (dashed), black and blue lines
    respectively.}
  \label{fig:1Dproject}
\end{figure}

\begin{figure}
  \centering
  \includegraphics[width=.8\textwidth]{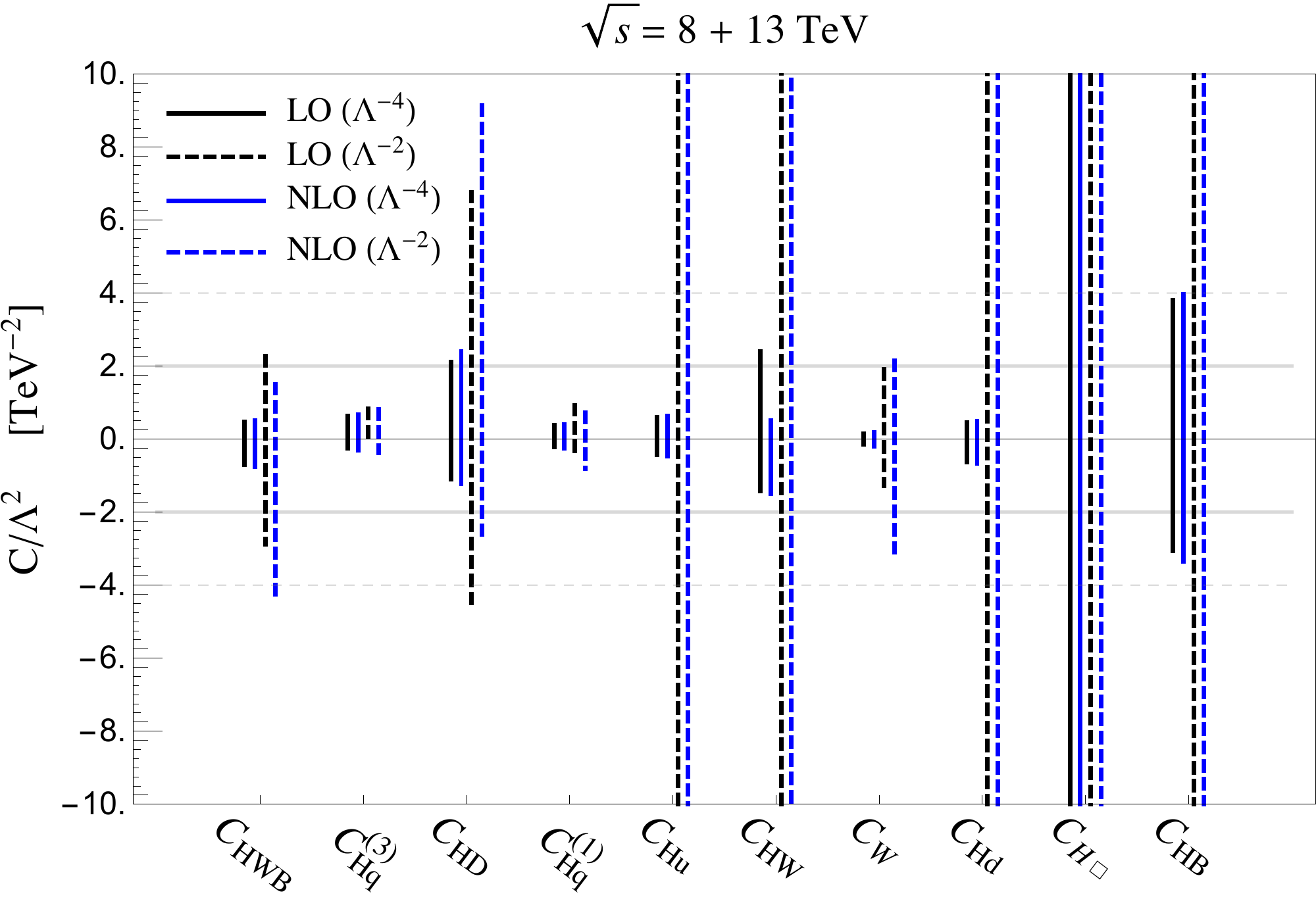}
  \caption{$95 \%$ confidence interval fits to individual EFT
    coefficients using $W^{+}W^{-}+W^{\pm}Z + ZH + W^{\pm}H$ data and
    profiling over all other coefficients. Fits quadratic in
    $1/\Lambda^2$  (Linear) LO and NLO QCD are shown as solid (dashed)
    black and blue lines respectively.}
  \label{fig:1Dprofile}
\end{figure}

The $95 \%$ confidence intervals for the projected individual
parameters are shown in Figure~\ref{fig:1Dproject}. We have included
solid (dashed) grey lines at $\pm 0.5 (1.0)$ to guide the
eye. Similarly, we show the individual $95 \%$ confidence intervals
from the profiled fitting procedure in Figure~\ref{fig:1Dprofile}. The
solid (dashed) lines are now at $\pm 2.0 (4.0)$ and the scales have
been expanded.  Black (blue) lines indicate that we are working to LO
(NLO) QCD in the SMEFT, and solid (dashed) lines indicate the
expansion to  $\mathcal{O}(\frac{1}{\Lambda^4})$ (
$\mathcal{O}(\frac{1}{\Lambda^2})$ ).

\begin{figure}[b]
  \centering
  \includegraphics[width=.9\textwidth]{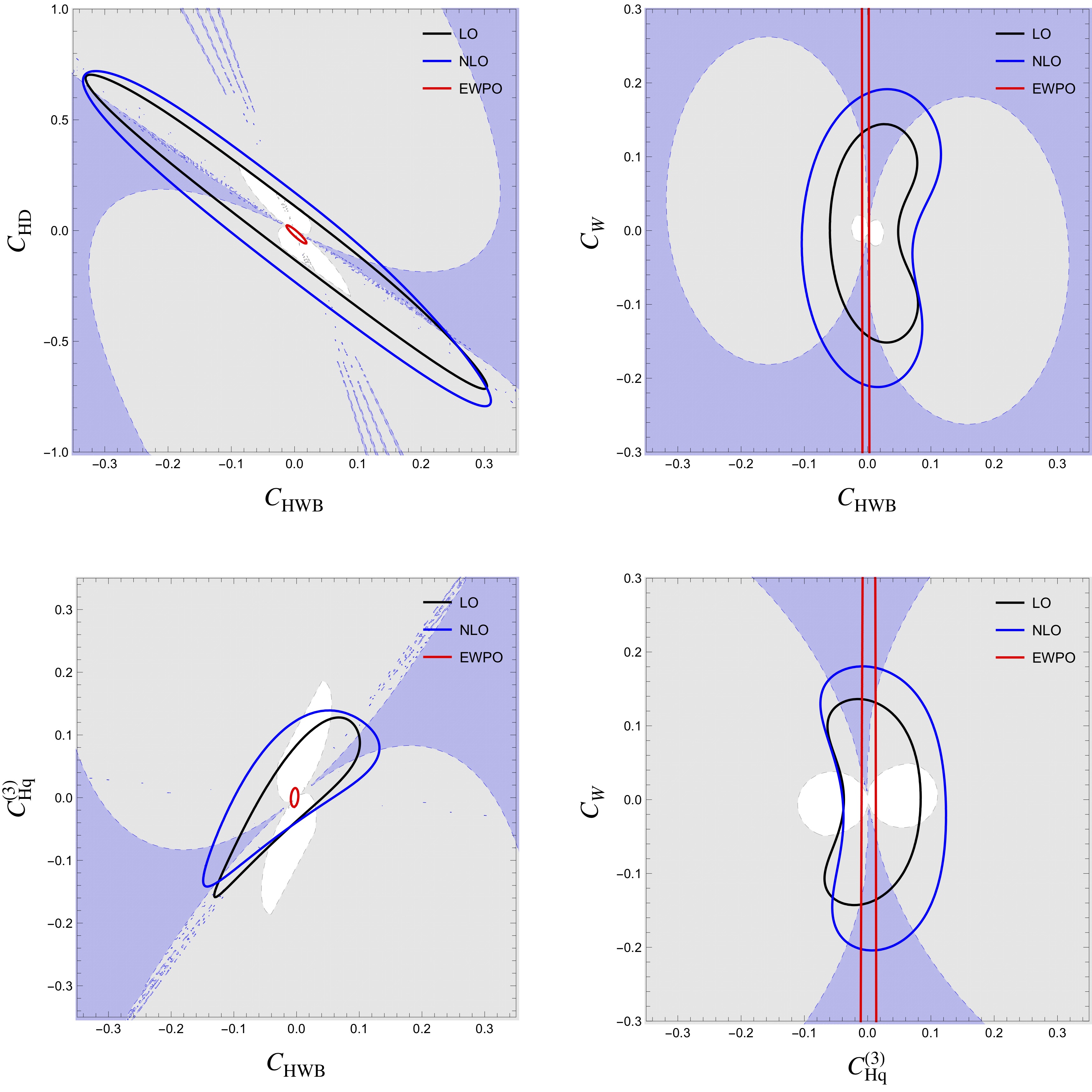}
  \caption{95 \% confidence region in the $C_{HWB}-C_{HD}$ (top
    left) , $C_{HWB}-C_{W}$  (top right), $C_{HWB}-C_{Hq}^{(3)}$
    (bottom left), and $C_{Hq}^{(3)}-C_{W}$ (bottom right)
    planes with all other EFT coefficients projected to zero and
    $\Lambda$ fixed to $1~\mathrm{TeV}$. Quadratic fits to
    $W^{+}W^{-}+W^{\pm}Z + ZH + W^{\pm}H$ distributions are
    shown in black (blue) for LO (NLO) QCD in SMEFT, while the
    quadratic fit to NLO Electroweak Precision Observables is
    shown in red. The grey (blue) region indicates the
    coefficients no longer correspond to a weakly-coupled theory
    that is $\Delta \sigma_{\Lambda^4} > \Delta
    \sigma_{\Lambda^2} ~(4 \pi  \Delta \sigma_{\Lambda^2}) $. }
  \label{fig:2Dplots}
\end{figure}

Similarly,  we show the $ 95 \%$ confidence intervals for some
selected planes of parameters using the projected method in
Figure~\ref{fig:2Dplots} for LO (inside black curve ) and NLO (inside
blue curve) QCD in the SMEFT, along with the limits from Electroweak
Precision Observables (EWPO)~\cite{Falkowski:2014tna} (inside red
curve) to $\mathcal{O}(\frac{1}{\Lambda^4})$, using the $\chi^2$ fit
of Ref.~\cite{Dawson:2019clf} \footnote{Ref.~\cite{Dawson:2019clf}
  demonstrates in the case of the EWPO the important effects from
  including both QCD and electroweak SMEFT NLO corrections.}. Again,
we emphasize that our results are not meant to compete with the global
fits including Higgs data and EWPO, but rather, our goal is to
determine the importance of NLO QCD within the SMEFT and to examine
the $1/\Lambda^2$ dependence. The EWPO curves are included, however,
as a reference for comparison.

First, let us compare the differences of the LO and NLO QCD  fits in
the SMEFT, the black and blue lines. Looking at the results in
Figures~\ref{fig:1Dproject} and \ref{fig:1Dprofile}, including NLO QCD
in the SMEFT can change the fit intervals on the order of $10-20 \%$,
on average. For some coefficients,  NLO QCD can have an effect as
large as $50 \%$. From the two-dimensional plots in
Figure~\ref{fig:2Dplots},  we  see that going from LO to NLO QCD can
shift the curves by  as much as $25 \%$ in some directions, along
with altering  the overall orientation and shapes of  the curves.

Next we  compare the differences in the fits when working to
$\mathcal{O}(\frac{1}{\Lambda^4})$ versus
$\mathcal{O}(\frac{1}{\Lambda^2})$, the solid and dashed lines. The
$\mathcal{O}(\frac{1}{\Lambda^4})$ fits are always better or
comparable to the $\mathcal{O}(\frac{1}{\Lambda^2})$ fits.  On
average,  working to  $\mathcal{O}(\frac{1}{\Lambda^4})$ improves the
fits by a factor of two and as much as  a factor of ten in some of the
profiled fits. Such a large improvement in the fit hints that the
coefficients no longer correspond to a weakly-coupled theory and we
discuss this in more detail in the following section. Similar results
when comparing the   $\mathcal{O}(\frac{1}{\Lambda^4})$ fits to those
obtained at $\mathcal{O}(\frac{1}{\Lambda^2})$ were obtained in
Ref.~\cite{Almeida:2018cld}  at LO QCD.

\subsection{The Validity of Weakly-Coupled Theory}

We decompose the differential  cross sections as in  Eq.~\ref{eq:dec} .
The SMEFT couplings generically scale as $\alpha_{\text{EFT}} \sim
\frac{g_{\text{EFT}}^2 v^2 }{\Lambda^2}$ or $ \frac{g_{\text{EFT}}^2
  {\text{Energy}}^2 }{\Lambda^2}$, where $g_{EFT}$ parameterizes the
strength of the underlying  UV  complete theory.  The linear piece,
$\Delta \sigma_{\Lambda^2}  $, goes as  $\mathcal{O}
(\alpha_{\text{EFT}})$, and the quadratic piece $\Delta
\sigma_{\Lambda^4} $ goes as $\mathcal{O} (\alpha_{\text{EFT}}^2)$. In
a weakly-coupled theory, one generically expects $\alpha_{\text{EFT}}
\alt 1 $.  This implies  $ \Delta \sigma_{\Lambda^4}  / \Delta
\sigma_{\Lambda^2}  \alt 1 $ for a weakly-coupled theory, assuming
that there are no cancellations in the underlying UV
theory. Alternatively, one might also consider the upper limit on a
weakly-coupled theory to be $\alpha_{\text{EFT}} \alt 4 \pi $ as some
sort of perturbative unitarity bound.  Similar criterion have been
explored elsewhere in the literature~\cite{Contino:2016jqw,}.

In Fig.~\ref{fig:2Dplots}, we show different regions detailing the
strength of the coupling by comparing the differential cross sections
in each bin. All parameters not shown in the plot are projected to
zero. The white regions in the figures indicate that $ \Delta
\sigma_{\Lambda^4}  / \Delta \sigma_{\Lambda^2}  < 1 $ for all bins in
all  of the processes considered. One may consider this the
weakly-coupled regime. The grey and blue regions respectively indicate
that $\Delta \sigma_{\Lambda^4}  / \Delta \sigma_{\Lambda^2}  >1 $ and
$ \Delta \sigma_{\Lambda^4}  / \Delta \sigma_{\Lambda^2}  > 4 \pi $ in
at least one bin for at least one process.  Any coefficient or fit in
these regions would no longer be considered part of a weakly-coupled
theory.

We see in Fig.~\ref{fig:2Dplots} that many of the confidence intervals
we derived for the $WV + VH$ data  (within the blue or black curves)
fall within a grey shaded region. If the coefficients lie in this area
they correspond to a strongly-interacting theory and higher-dimension
operators need to be retained.  In contrast, the bounds from the EWPO
(within the red curves) place strong constraints on the couplings and
typically fall within the weakly-coupled regime (white region). One
might consider setting an experimental goal of LHC to have all fits
sufficiently precise such that they could probe the weakly-coupled
regime. In this way one could fully understand the fits in terms of
dimension-6 operators.

There are small regions protruding into some of the regions within the
plots. They are particularly evident in the top left plot in
Figure~\ref{fig:2Dplots}. These can be seen in other plots not
displayed here. They can be understood as cancellations within the
helicity amplitudes.

We also note that as Warsaw coefficients are increased, the last bin
will be the first indication that the weakly-coupled theory is no
longer valid. The argument is similar to those made in previous works
showing that most of the fitting power comes from the last
bin~\cite{Brehmer:2019gmn}. We know SM cross sections are falling with
increasing energy, while the quadratic SMEFT piece grows with energy.
Therefore the bin with the largest energy, the last one, will have the
largest deviation from the SM and best fitting power.

\section{Conclusion}
\label{sec:conc}

As the quest for discovering beyond-Standard-Model particles continues
without any direct observations, it is important to understand all the
data we have to the best precision.  Such precision measurements could
be the first evidence for some new high-scale physics. To this end, we
have studied the effects of NLO QCD in the SMEFT on the $W^+W^-,
W^\pm Z, W^\pm H$, and $ZH$ production at the LHC. We find that
including QCD radiation can have a significant effect on the
parameters. This implies that global SMEFT fits including Higgs data
and EWPO need to be done beyond LO in QCD. We have also explored the
numerical differences between the $1/\Lambda^2$ and $1/\Lambda^4$
fits. Their differences suggest that current fits to LHC data are not
yet sensitive to weakly-coupled theories for the majority of
coefficients.

Primitive cross sections at  $8$ and $13$~TeV for $W^\pm Z$ production
with jet vetos, and at  $13$~TeV for $W^\pm H$ production are posted
at
\url{https://quark.phy.bnl.gov/Digital_Data_Archive/dawson/VV_20}.

\begin{acknowledgments}
SD is supported by the United States Department of Energy under Grant
Contract DE-SC0012704.
The work of SH was supported in part by the National Science
Foundation grant PHY-1915093.
IML is supported in part by United States Department of Energy grant
number DE-SC0017988. 
SL is supported by the State of Kansas EPSCoR grant program and the
U.S. Department of Energy, Office of Science, Office of Workforce
Development for Teachers and Scientists, Office of Science Graduate
Student Research (SCGSR) program.
The SCGSR program is administered by the Oak Ridge Institute for
Science and Education (ORISE) for the DOE.
ORISE is managed by ORAU under contract number DE-SC0014664.
\end{acknowledgments}

\appendix
\section{Numerical Fits}

We show tables detailing the numerical results of the $95 \%$
confidence intervals to different subsets of processes in
Tables~\ref{tab:WHZHfits}, ~\ref{tab:WWWZfits}, and
~\ref{tab:WVVHfits}. Entries with a "-" mean no fit was performed,
since the process does not depend on that parameter.  Overall, fitting
to a few bins in $W^\pm H$ and $ZH$ processes yields comparable
sensitivity to that of the $W^+W^- $and $W^\pm Z$ fits for some
parameters.

\begin{table}[!h]
\center
\begin{tabular}{|c|c|c|c|c|c|c|c|c|c|c|c|c|}
\hline
 \text{} & \multicolumn{4}{c|}{$W^{\pm}H$ Projected}&   \multicolumn{4}{c|}{$ZH$ Projected} \\
 \cline{2-9}
 & \multicolumn{2}{c|}{$\Lambda^{-4}$} & \multicolumn{2}{c|}{$\Lambda^{-2}$} & 
     \multicolumn{2}{c|}{$\Lambda^{-4}$} & \multicolumn{2}{c|}{$\Lambda^{-2}$} \\
 \cline{2-9}
  & \text{LO }& \text{NLO }& \text{LO } & \text{NLO } 
  & \text{LO }& \text{NLO}& \text{LO }& \text{NLO }\\
\hline 
$ C_{\text{HWB}} $
& - & - & - & - 
& (-3.1, 1.8) & (-3.3, 1.8) & (-1.8, 3.4) & (-1.8, 3.4) \\
$ C_{\text{Hq}}^{(3)} $
& (-.61, .19) & (-.65, .20) & (-.20, .26) & (-.22, .29) 
& (-.33, .12) & (-.35, .13) & (-.08, .18) & (-.09, .20)  \\
$ C_{\text{HD}} $
& (-33., 16.) & (-33., 16.) & (-63., 53.) & (-63., 53.) 
& (-17., 21.) & (-17., 21.) & (-14., 26.) & (-14., 26.)  \\
$ C_{\text{Hq}}^{(1)}$ 
& - & - & - & - 
& (-.20, .22) & (-.21, .24) & (-1.9, .77) & (-2.0, .82)  \\
$ C_{\text{Hu}} $ 
& - & - & - & - 
& (-.31, .22) & (-.34, .24) & (-.33, .76) & (-.37, .85) \\
$ C_{\text{HW}} $ 
& (-1.2, .59) & (-1.2, .61) & (-.96, 1.1) & (-.96, 1.1) 
& (-1.5, .75) & (-1.5, .77) & (-.69, 1.3) & (-.69, 1.3) \\
$ C_W $ 
& - & - & - & - 
& - & - & - & - \\
$ C_{\text{Hd}} $ 
& - & - & - & - 
& (-.31, .36) & (-.34, .39) & (-2.3, 1.) & (-2.3, 1.0)  \\
$ C_{H\square }$ 
& (-41., 8.2) & (-41., 8.2) & (-13., 16.) & (-13., 16.) 
& (-6.5, 7.8) & (-6.5, 7.8) & (-5.2, 9.7) & (-5.2, 9.7) \\
$ C_{\text{HB}} $ 
& - & - & - & - 
& (-2.8, 2.3) & (-2.9, 2.4) & (-6.1, 11.) & (-6.0, 12.)  \\
\hline
\end{tabular}
\caption{$95\%$ confidence interval fits to individual EFT
  coefficients using $W^\pm H$ and $ZH$ data, with $\Lambda$ fixed to
  $1~\mathrm{TeV}$.}
\label{tab:WHZHfits}
\end{table}

\begin{table}[!h]
\begin{tabular}{|c|c|c|c|c|c|c|c|c|}
\hline
 \text{} & \multicolumn{4}{c|}{$W^+W^-$ Projected}&   \multicolumn{4}{c|}{$W^{\pm}Z$ Projected} \\
 \cline{2-9}
 & \multicolumn{2}{c|}{$\Lambda^{-4}$} & \multicolumn{2}{c|}{$\Lambda^{-2}$} & 
     \multicolumn{2}{c|}{$\Lambda^{-4}$} & \multicolumn{2}{c|}{$\Lambda^{-2}$} \\
 \cline{2-9}
  & \text{LO }& \text{NLO }& \text{LO } & \text{NLO } 
  & \text{LO }& \text{NLO}& \text{LO }& \text{NLO }\\
\hline 
 $C_{\text{HWB}}$ 
& (-.14, .17) & (-.14, .18) & (-.35, .38) & (-.37, .4) 
& (-.05, .03) & (-.1, .03) & (-.07, .02) & (-.14, .03) \\
$ C_{\text{Hq}}^{(3)}$ 
& (-.34, .21) & (-.35, .22) & (-.33, .3) & (-.35, .32) 
& (-.03, .08) & (-.03, .15) & (-.03, .1) & (-.03, .18)  \\
$ C_{\text{HD}} $
& (-.35, .54) & (-.36, .56) & (-.60, .69) & (-.64, .73) 
& (-.12, .06) & (-.22, .07) & (-.15, .05) & (-.32, .06) \\
$C_{\text{Hq}}^{(1)} $
& (-.37, .34) & (-.38, .35) & (-4.8, 3.1) & (-5.4, 3.4) 
& (-.26, 1.7) & (-1.5, .43) & (-.15, 2.8) & (-1.3, .45) \\
$ C_{\text{Hu}} $
& (-.47, .41) & (-.48, .42) & (-3.1, 2.4) & (-3.4, 2.6) 
& - & - & - & -   \\
 $C_{\text{HW}} $ 
& - & - & - & - 
& - & - & - & - \\
$ C_W $
& (-.22, .23) & (-.23, .23) & (-4.4, 5.1) & (-9.6, 6.8) 
& (-.14, .13) & (-.22, .19) & (-1.5, 1.3) & (-.27, .94)  \\
 $C_{\text{Hd}} $
& (-.59, .62) & (-.59, .63) & (-7.6, 9.7) & (-8.0, 10.) 
& - & - & - & -   \\
 $C_{H\square } $
 & - & - & - & - 
 & - & - & - & -  \\
 $C_{\text{HB}}$ 
 & - & - & - & - 
 & - & - & - & -  \\
 \hline
\end{tabular}
\caption{The same as Table~\ref{tab:WHZHfits}, but using $W^+W^-$ and
  $W^{\pm}Z$ data, with $\Lambda$ fixed to $1~\mathrm{TeV}$}
\label{tab:WWWZfits}
\end{table}

\begin{table}[!h]
\begin{tabular}{|c|c|c|c|c|c|c|c|c|}
\hline
 \text{}& \multicolumn{4}{c|}{$ZH + W^{\pm}H$ Projected} &   \multicolumn{4}{c|}{$W^{+}W^{-}+W^{\pm}Z $ Projected} \\
 \cline{2-9}
 & \multicolumn{2}{c|}{$\Lambda^{-4}$} & \multicolumn{2}{c|}{$\Lambda^{-2}$} & 
     \multicolumn{2}{c|}{$\Lambda^{-4}$} & \multicolumn{2}{c|}{$\Lambda^{-2}$} \\
 \cline{2-9}
  & \text{LO }& \text{NLO }& \text{LO } & \text{NLO } 
  & \text{LO }& \text{NLO}& \text{LO }& \text{NLO }\\
\hline 
$C_{\text{HWB}}$ 
& (-3.1, 1.8) & (-3.3, 1.8) & (-1.8, 3.4) & (-1.8, 3.4)
& (-.05, .03) & (-.09, .04) & (-.07, .02) & (-.14, .03) \\
$ C_{\text{Hq}}^{(3)}$ 
& (-.32, .12) & (-.34, .13) & (-.07, .16) & (-.07, .18)
& (-.03, .08) & (-.03, .14) & (-.03, .10) & (-.03, .17) \\
$ C_{\text{HD}} $
& (-16., 19.) & (-16., 19.) & (-14., 24.) & (-14., 24.)
& (-.12, .06) & (-.21, .08) & (-.15, .05) & (-.30, .07) \\
$C_{\text{Hq}}^{(1)} $
& (-.17, .21) & (-.18, .23) & (-.27, .18) & (-.30, .20)
& (-.31, .37) & (-.40, .28) & (-.33, 2.5) & (-1.3, .41) \\
$ C_{\text{Hu}} $
& (-.31, .22) & (-.34, .24) & (-.33, .76) & (-.37, .85)
& (-.47, .41) & (-.48, .42) & (-3.1, 2.4) & (-3.4, 2.6) \\
$C_{\text{HW}} $
& (-1.1, .55) & (-1.2, .56) & (-.52, .92) & (-.52, .92)
& - & - & - & - \\
$ C_W $
& - & - & - & -
& (-.13, .13) & (-.20, .18) & (-1.4, 1.3) & (-.28, .93) \\
$C_{\text{Hd}} $
& (-.31, .36) & (-.34, .39) & (-2.3, 1.0) & (-2.3, 1.0)
& (-.59, .62) & (-.59, .63) & (-7.6, 9.7) & (-8.0, 10.) \\
$C_{H\square } $
& (-4.9, 6.3) & (-4.9, 6.3) & (-4.6, 8.6) & (-4.6, 8.6)
& - & - & - & - \\
$C_{\text{HB}}$
& (-2.8, 2.3) & (-2.9, 2.4) & (-6.1, 11.) & (-6.0, 12.)
& - & - & - & - \\
\hline
\end{tabular}
\caption{The same as Table~\ref{tab:WHZHfits}, but using $ZH +
  W^{\pm}H$ and $W^{+}W^{-}+W^{\pm}Z $ data, with $\Lambda$ fixed to
  $1~\mathrm{TeV}$.}
\label{tab:WVVHfits}
\end{table}

\clearpage
\bibliographystyle{utphys}
\bibliography{vh.bib}

\providecommand{\href}[2]{#2}\begingroup\raggedright\begin{thebibliography}{10}

\bibitem{Almeida:2018cld}
E.~{da Silva Almeida}, A.~Alves, N.~Rosa~Agostinho, O.~J.~P. \'{E}boli, and
  M.~C. Gonzalez-Garcia, ``{Electroweak Sector Under Scrutiny: A Combined
  Analysis of LHC and Electroweak Precision Data},''
  \href{http://dx.doi.org/10.1103/PhysRevD.99.033001}{{\em Phys. Rev.}
  {\bfseries D99} no.~3, (2019) 033001},
\href{http://arxiv.org/abs/1812.01009}{{\ttfamily arXiv:1812.01009 [hep-ph]}}.

\bibitem{Biekotter:2018rhp}
A.~Biekotter, T.~Corbett, and T.~Plehn, ``{The Gauge-Higgs Legacy of the LHC
  Run II},'' \href{http://dx.doi.org/10.21468/SciPostPhys.6.6.064}{{\em SciPost
  Phys.} {\bfseries 6} (2019) 064},
\href{http://arxiv.org/abs/1812.07587}{{\ttfamily arXiv:1812.07587 [hep-ph]}}.

\bibitem{Grojean:2018dqj}
C.~Grojean, M.~Montull, and M.~Riembau, ``{Diboson at the LHC vs LEP},''
  \href{http://dx.doi.org/10.1007/JHEP03(2019)020}{{\em JHEP} {\bfseries 03}
  (2019) 020},
\href{http://arxiv.org/abs/1810.05149}{{\ttfamily arXiv:1810.05149 [hep-ph]}}.

\bibitem{Ellis:2018gqa}
J.~Ellis, C.~W. Murphy, V.~Sanz, and T.~You, ``{Updated Global SMEFT Fit to
  Higgs, Diboson and Electroweak Data},''
  \href{http://dx.doi.org/10.1007/JHEP06(2018)146}{{\em JHEP} {\bfseries 06}
  (2018) 146},
\href{http://arxiv.org/abs/1803.03252}{{\ttfamily arXiv:1803.03252 [hep-ph]}}.

\bibitem{Berthier:2016tkq}
L.~Berthier, M.~Bjorn, and M.~Trott, ``{Incorporating doubly resonant $W^\pm$
  data in a global fit of SMEFT parameters to lift flat directions},''
  \href{http://dx.doi.org/10.1007/JHEP09(2016)157}{{\em JHEP} {\bfseries 09}
  (2016) 157},
\href{http://arxiv.org/abs/1606.06693}{{\ttfamily arXiv:1606.06693 [hep-ph]}}.

\bibitem{Brivio:2017vri}
I.~Brivio and M.~Trott, ``{The Standard Model as an Effective Field Theory},''
  \href{http://dx.doi.org/10.1016/j.physrep.2018.11.002}{{\em Phys. Rept.}
  {\bfseries 793} (2019) 1--98},
\href{http://arxiv.org/abs/1706.08945}{{\ttfamily arXiv:1706.08945 [hep-ph]}}.

\bibitem{}
A.~Falkowski, M.~Gonzalez-Alonso, A.~Greljo, D.~Marzocca, and M.~Son,
  ``{Anomalous Triple Gauge Couplings in the Effective Field Theory Approach at
  the LHC},'' \href{http://dx.doi.org/10.1007/JHEP02(2017)115}{{\em JHEP}
  {\bfseries 02} (2017) 115},
\href{http://arxiv.org/abs/1609.06312}{{\ttfamily arXiv:1609.06312 [hep-ph]}}.

\bibitem{Buchmuller:1985jz}
W.~Buchmuller and D.~Wyler, ``{Effective Lagrangian Analysis of New
  Interactions and Flavor Conservation},''
\href{http://dx.doi.org/10.1016/0550-3213(86)90262-2}{{\em Nucl. Phys.}
  {\bfseries B268} (1986) 621--653}.

\bibitem{Grzadkowski:2010es}
B.~Grzadkowski, M.~Iskrzynski, M.~Misiak, and J.~Rosiek, ``{Dimension-Six Terms
  in the Standard Model Lagrangian},''
  \href{http://dx.doi.org/10.1007/JHEP10(2010)085}{{\em JHEP} {\bfseries 10}
  (2010) 085},
\href{http://arxiv.org/abs/1008.4884}{{\ttfamily arXiv:1008.4884 [hep-ph]}}.

\bibitem{Baglio:2017bfe}
J.~Baglio, S.~Dawson, and I.~M. Lewis, ``{An NLO QCD effective field theory
  analysis of $W^+W^-$ production at the LHC including fermionic operators},''
  \href{http://dx.doi.org/10.1103/PhysRevD.96.073003}{{\em Phys. Rev.}
  {\bfseries D96} no.~7, (2017) 073003},
\href{http://arxiv.org/abs/1708.03332}{{\ttfamily arXiv:1708.03332 [hep-ph]}}.

\bibitem{Baglio:2018bkm}
J.~Baglio, S.~Dawson, and I.~M. Lewis, ``{NLO effects in EFT fits to $W^+W^-$
  production at the LHC},''
  \href{http://dx.doi.org/10.1103/PhysRevD.99.035029}{{\em Phys. Rev.}
  {\bfseries D99} no.~3, (2019) 035029},
\href{http://arxiv.org/abs/1812.00214}{{\ttfamily arXiv:1812.00214 [hep-ph]}}.

\bibitem{Baglio:2019uty}
J.~Baglio, S.~Dawson, and S.~Homiller, ``{QCD corrections in Standard Model EFT
  fits to $WZ$ and $WW$ production},''
  \href{http://dx.doi.org/10.1103/PhysRevD.100.113010}{{\em Phys. Rev.}
  {\bfseries D100} no.~11, (2019) 113010},
\href{http://arxiv.org/abs/1909.11576}{{\ttfamily arXiv:1909.11576 [hep-ph]}}.

\bibitem{Alioli:2018ljm}
S.~Alioli, W.~Dekens, M.~Girard, and E.~Mereghetti, ``{NLO QCD corrections to
  SM-EFT dilepton and electroweak Higgs boson production, matched to parton
  shower in POWHEG},'' \href{http://dx.doi.org/10.1007/JHEP08(2018)205}{{\em
  JHEP} {\bfseries 08} (2018) 205},
\href{http://arxiv.org/abs/1804.07407}{{\ttfamily arXiv:1804.07407 [hep-ph]}}.

\bibitem{Melia:2011tj}
T.~Melia, P.~Nason, R.~Rontsch, and G.~Zanderighi, ``{W+W-, WZ and ZZ
  production in the POWHEG BOX},''
  \href{http://dx.doi.org/10.1007/JHEP11(2011)078}{{\em JHEP} {\bfseries 11}
  (2011) 078},
\href{http://arxiv.org/abs/1107.5051}{{\ttfamily arXiv:1107.5051 [hep-ph]}}.

\bibitem{Nason:2013ydw}
P.~Nason and G.~Zanderighi, ``{$W^+ W^-$ , $W Z$ and $Z Z$ production in the
  POWHEG-BOX-V2},''
  \href{http://dx.doi.org/10.1140/epjc/s10052-013-2702-5}{{\em Eur. Phys. J.}
  {\bfseries C74} no.~1, (2014) 2702},
\href{http://arxiv.org/abs/1311.1365}{{\ttfamily arXiv:1311.1365 [hep-ph]}}.

\bibitem{Luisoni:2013kna}
G.~Luisoni, P.~Nason, C.~Oleari, and F.~Tramontano, ``{$HW^{\pm}$/HZ + 0 and 1
  jet at NLO with the POWHEG BOX interfaced to GoSam and their merging within
  MiNLO},'' \href{http://dx.doi.org/10.1007/JHEP10(2013)083}{{\em JHEP}
  {\bfseries 10} (2013) 083},
\href{http://arxiv.org/abs/1306.2542}{{\ttfamily arXiv:1306.2542 [hep-ph]}}.

\bibitem{Frixione:2007vw}
S.~Frixione, P.~Nason, and C.~Oleari, ``{Matching NLO QCD computations with
  Parton Shower simulations: the POWHEG method},''
  \href{http://dx.doi.org/10.1088/1126-6708/2007/11/070}{{\em JHEP} {\bfseries
  11} (2007) 070},
\href{http://arxiv.org/abs/0709.2092}{{\ttfamily arXiv:0709.2092 [hep-ph]}}.

\bibitem{Alioli:2010xd}
S.~Alioli, P.~Nason, C.~Oleari, and E.~Re, ``{A general framework for
  implementing NLO calculations in shower Monte Carlo programs: the POWHEG
  BOX},'' \href{http://dx.doi.org/10.1007/JHEP06(2010)043}{{\em JHEP}
  {\bfseries 06} (2010) 043},
\href{http://arxiv.org/abs/1002.2581}{{\ttfamily arXiv:1002.2581 [hep-ph]}}.

\bibitem{Ohnemus:1991kk}
J.~Ohnemus, ``{An Order $\alpha_s$ calculation of hadronic $W^{-} W^{+}$
  production},''
\href{http://dx.doi.org/10.1103/PhysRevD.44.1403}{{\em Phys. Rev.} {\bfseries
  D44} (1991) 1403--1414}.

\bibitem{Ohnemus:1991gb}
J.~Ohnemus, ``{An Order $\alpha_s$ calculation of hadronic $W^\pm Z$
  production},''
\href{http://dx.doi.org/10.1103/PhysRevD.44.3477}{{\em Phys. Rev.} {\bfseries
  D44} (1991) 3477--3489}.

\bibitem{Frixione:1992pj}
S.~Frixione, P.~Nason, and G.~Ridolfi, ``{Strong corrections to W Z production
  at hadron colliders},''
\href{http://dx.doi.org/10.1016/0550-3213(92)90668-2}{{\em Nucl. Phys.}
  {\bfseries B383} (1992) 3--44}.

\bibitem{Ohnemus:1994ff}
J.~Ohnemus, ``{Hadronic $Z Z$, $W^{-} W^{+}$, and $W^\pm Z$ production with QCD
  corrections and leptonic decays},''
  \href{http://dx.doi.org/10.1103/PhysRevD.50.1931}{{\em Phys. Rev.} {\bfseries
  D50} (1994) 1931--1945},
\href{http://arxiv.org/abs/hep-ph/9403331}{{\ttfamily arXiv:hep-ph/9403331
  [hep-ph]}}.

\bibitem{Dixon:1998py}
L.~J. Dixon, Z.~Kunszt, and A.~Signer, ``{Helicity amplitudes for O(alpha-s)
  production of $W^{+} W^{-}$, $W^\pm Z$, $Z Z$, $W^\pm \gamma$, or $Z \gamma$
  pairs at hadron colliders},''
  \href{http://dx.doi.org/10.1016/S0550-3213(98)00421-0}{{\em Nucl. Phys.}
  {\bfseries B531} (1998) 3--23},
\href{http://arxiv.org/abs/hep-ph/9803250}{{\ttfamily arXiv:hep-ph/9803250
  [hep-ph]}}.

\bibitem{Campbell:1999ah}
J.~M. Campbell and R.~K. Ellis, ``{An Update on vector boson pair production at
  hadron colliders},'' \href{http://dx.doi.org/10.1103/PhysRevD.60.113006}{{\em
  Phys. Rev.} {\bfseries D60} (1999) 113006},
\href{http://arxiv.org/abs/hep-ph/9905386}{{\ttfamily arXiv:hep-ph/9905386
  [hep-ph]}}.

\bibitem{Campbell:2011bn}
J.~M. Campbell, R.~K. Ellis, and C.~Williams, ``{Vector boson pair production
  at the LHC},'' \href{http://dx.doi.org/10.1007/JHEP07(2011)018}{{\em JHEP}
  {\bfseries 07} (2011) 018},
\href{http://arxiv.org/abs/1105.0020}{{\ttfamily arXiv:1105.0020 [hep-ph]}}.

\bibitem{Han:1991ia}
T.~Han and S.~Willenbrock, ``{QCD correction to the $p p \to W H$ and $Z H$
  total cross-sections},''
\href{http://dx.doi.org/10.1016/0370-2693(91)90572-8}{{\em Phys. Lett.}
  {\bfseries B273} (1991) 167--172}.

\bibitem{Ohnemus:1992bd}
J.~Ohnemus and W.~J. Stirling, ``{Order $\alpha_s$ corrections to the
  differential cross-section for the $W H$ intermediate mass Higgs signal},''
\href{http://dx.doi.org/10.1103/PhysRevD.47.2722}{{\em Phys. Rev.} {\bfseries
  D47} (1993) 2722--2729}.

\bibitem{Baer:1992vx}
H.~Baer, B.~Bailey, and J.~F. Owens, ``{$\mathcal{O}(\alpha_s)$ Monte Carlo
  approach to W + Higgs associated production at hadron supercolliders},''
\href{http://dx.doi.org/10.1103/PhysRevD.47.2730}{{\em Phys. Rev.} {\bfseries
  D47} (1993) 2730--2734}.

\bibitem{Stange:1994bb}
A.~Stange, W.~J. Marciano, and S.~Willenbrock, ``{Associated production of
  Higgs and weak bosons, with $H\to b\bar{b}$, at hadron colliders},''
  \href{http://dx.doi.org/10.1103/PhysRevD.50.4491}{{\em Phys. Rev.} {\bfseries
  D50} (1994) 4491--4498},
\href{http://arxiv.org/abs/hep-ph/9404247}{{\ttfamily arXiv:hep-ph/9404247
  [hep-ph]}}.

\bibitem{Gehrmann:2014fva}
T.~Gehrmann, M.~Grazzini, S.~Kallweit, P.~Maierhofer, A.~von Manteuffel,
  S.~Pozzorini, D.~Rathlev, and L.~Tancredi, ``{$W^+W^-$ Production at Hadron
  Colliders in Next to Next to Leading Order QCD},''
  \href{http://dx.doi.org/10.1103/PhysRevLett.113.212001}{{\em Phys. Rev.
  Lett.} {\bfseries 113} no.~21, (2014) 212001},
\href{http://arxiv.org/abs/1408.5243}{{\ttfamily arXiv:1408.5243 [hep-ph]}}.

\bibitem{Caola:2015rqy}
F.~Caola, K.~Melnikov, R.~Rontsch, and L.~Tancredi, ``{QCD corrections to
  $W^+W^-$ production through gluon fusion},''
  \href{http://dx.doi.org/10.1016/j.physletb.2016.01.046}{{\em Phys. Lett.}
  {\bfseries B754} (2016) 275--280},
\href{http://arxiv.org/abs/1511.08617}{{\ttfamily arXiv:1511.08617 [hep-ph]}}.

\bibitem{Grazzini:2016swo}
M.~Grazzini, S.~Kallweit, D.~Rathlev, and M.~Wiesemann, ``{$W^{\pm}Z$
  production at hadron colliders in NNLO QCD},''
  \href{http://dx.doi.org/10.1016/j.physletb.2016.08.017}{{\em Phys. Lett.}
  {\bfseries B761} (2016) 179--183},
\href{http://arxiv.org/abs/1604.08576}{{\ttfamily arXiv:1604.08576 [hep-ph]}}.

\bibitem{Brein:2003wg}
O.~Brein, A.~Djouadi, and R.~Harlander, ``{NNLO QCD corrections to the
  Higgs-strahlung processes at hadron colliders},''
  \href{http://dx.doi.org/10.1016/j.physletb.2003.10.112}{{\em Phys. Lett.}
  {\bfseries B579} (2004) 149--156},
\href{http://arxiv.org/abs/hep-ph/0307206}{{\ttfamily arXiv:hep-ph/0307206
  [hep-ph]}}.

\bibitem{Ferrera:2011bk}
G.~Ferrera, M.~Grazzini, and F.~Tramontano, ``{Associated WH production at
  hadron colliders: a fully exclusive QCD calculation at NNLO},''
  \href{http://dx.doi.org/10.1103/PhysRevLett.107.152003}{{\em Phys. Rev.
  Lett.} {\bfseries 107} (2011) 152003},
\href{http://arxiv.org/abs/1107.1164}{{\ttfamily arXiv:1107.1164 [hep-ph]}}.

\bibitem{Ferrera:2014lca}
G.~Ferrera, M.~Grazzini, and F.~Tramontano, ``{Associated ZH production at
  hadron colliders: the fully differential NNLO QCD calculation},''
  \href{http://dx.doi.org/10.1016/j.physletb.2014.11.040}{{\em Phys. Lett.}
  {\bfseries B740} (2015) 51--55},
\href{http://arxiv.org/abs/1407.4747}{{\ttfamily arXiv:1407.4747 [hep-ph]}}.

\bibitem{Campbell:2016jau}
J.~M. Campbell, R.~K. Ellis, and C.~Williams, ``{Associated production of a
  Higgs boson at NNLO},'' \href{http://dx.doi.org/10.1007/JHEP06(2016)179}{{\em
  JHEP} {\bfseries 06} (2016) 179},
\href{http://arxiv.org/abs/1601.00658}{{\ttfamily arXiv:1601.00658 [hep-ph]}}.

\bibitem{Baglio:2013toa}
J.~Baglio, L.~D. Ninh, and M.~M. Weber, ``{Massive gauge boson pair production
  at the LHC: a next-to-leading order story},''
  \href{http://dx.doi.org/10.1103/PhysRevD.94.099902,
  10.1103/PhysRevD.88.113005}{{\em Phys. Rev.} {\bfseries D88} (2013) 113005},
  \href{http://arxiv.org/abs/1307.4331}{{\ttfamily arXiv:1307.4331 [hep-ph]}}.
[Erratum: Phys. Rev.D94,no.9,099902(2016)].

\bibitem{Bierweiler:2013dja}
A.~Bierweiler, T.~Kasprzik, and J.~H. Kuhn, ``{Vector-boson pair production at
  the LHC to $\mathcal{O}(\alpha^3)$ accuracy},''
  \href{http://dx.doi.org/10.1007/JHEP12(2013)071}{{\em JHEP} {\bfseries 12}
  (2013) 071},
\href{http://arxiv.org/abs/1305.5402}{{\ttfamily arXiv:1305.5402 [hep-ph]}}.

\bibitem{Billoni:2013aba}
M.~Billoni, S.~Dittmaier, B.~Jager, and C.~Speckner, ``{Next-to-leading order
  electroweak corrections to $pp \to W^+W^- \to$ 4 leptons at the LHC in
  double-pole approximation},''
  \href{http://dx.doi.org/10.1007/JHEP12(2013)043}{{\em JHEP} {\bfseries 12}
  (2013) 043},
\href{http://arxiv.org/abs/1310.1564}{{\ttfamily arXiv:1310.1564 [hep-ph]}}.

\bibitem{Biedermann:2016guo}
B.~Biedermann, M.~Billoni, A.~Denner, S.~Dittmaier, L.~Hofer, B.~J\"{a}ger, and
  L.~Salfelder, ``{Next-to-leading-order electroweak corrections to $pp \to
  W^+W^-\to$ 4 leptons at the LHC},''
  \href{http://dx.doi.org/10.1007/JHEP06(2016)065}{{\em JHEP} {\bfseries 06}
  (2016) 065},
\href{http://arxiv.org/abs/1605.03419}{{\ttfamily arXiv:1605.03419 [hep-ph]}}.

\bibitem{Biedermann:2017oae}
B.~Biedermann, A.~Denner, and L.~Hofer, ``{Next-to-leading-order electroweak
  corrections to the production of three charged leptons plus missing energy at
  the LHC},'' \href{http://dx.doi.org/10.1007/JHEP10(2017)043}{{\em JHEP}
  {\bfseries 10} (2017) 043},
\href{http://arxiv.org/abs/1708.06938}{{\ttfamily arXiv:1708.06938 [hep-ph]}}.

\bibitem{Baglio:2018rcu}
J.~Baglio and N.~Le~Duc, ``{Fiducial polarization observables in hadronic WZ
  production: A next-to-leading order QCD+EW study},''
  \href{http://dx.doi.org/10.1007/JHEP04(2019)065}{{\em JHEP} {\bfseries 04}
  (2019) 065},
\href{http://arxiv.org/abs/1810.11034}{{\ttfamily arXiv:1810.11034 [hep-ph]}}.

\bibitem{Kallweit:2019zez}
M.~Grazzini, S.~Kallweit, J.~M. Lindert, S.~Pozzorini, and M.~Wiesemann,
  ``{NNLO QCD + NLO EW with Matrix+OpenLoops: precise predictions for
  vector-boson pair production},''
  \href{http://dx.doi.org/10.1007/JHEP02(2020)087}{{\em JHEP} {\bfseries 02}
  (2020) 087},
\href{http://arxiv.org/abs/1912.00068}{{\ttfamily arXiv:1912.00068 [hep-ph]}}.

\bibitem{Denner:2011id}
A.~Denner, S.~Dittmaier, S.~Kallweit, and A.~Muck, ``{Electroweak corrections
  to Higgs-strahlung off W/Z bosons at the Tevatron and the LHC with HAWK},''
  \href{http://dx.doi.org/10.1007/JHEP03(2012)075}{{\em JHEP} {\bfseries 03}
  (2012) 075},
\href{http://arxiv.org/abs/1112.5142}{{\ttfamily arXiv:1112.5142 [hep-ph]}}.

\bibitem{Hagiwara:1986vm}
K.~Hagiwara, R.~D. Peccei, D.~Zeppenfeld, and K.~Hikasa, ``{Probing the Weak
  Boson Sector in $e^+ e^-\rightarrow W^+ W^-$},''
\href{http://dx.doi.org/10.1016/0550-3213(87)90685-7}{{\em Nucl. Phys.}
  {\bfseries B282} (1987) 253--307}.

\bibitem{Hagiwara:1993qt}
K.~Hagiwara, R.~Szalapski, and D.~Zeppenfeld, ``{Anomalous Higgs boson
  production and decay},''
  \href{http://dx.doi.org/10.1016/0370-2693(93)91799-S}{{\em Phys. Lett.}
  {\bfseries B318} (1993) 155--162},
\href{http://arxiv.org/abs/hep-ph/9308347}{{\ttfamily arXiv:hep-ph/9308347
  [hep-ph]}}.

\bibitem{Baur:1994aj}
U.~Baur, T.~Han, and J.~Ohnemus, ``{$W Z$ production at hadron colliders:
  Effects of nonstandard $W W Z$ couplings and QCD corrections},''
  \href{http://dx.doi.org/10.1103/PhysRevD.51.3381}{{\em Phys. Rev.} {\bfseries
  D51} (1995) 3381--3407},
\href{http://arxiv.org/abs/hep-ph/9410266}{{\ttfamily arXiv:hep-ph/9410266
  [hep-ph]}}.

\bibitem{Baur:1995uv}
U.~Baur, T.~Han, and J.~Ohnemus, ``{QCD corrections and nonstandard three
  vector boson couplings in $W^{+} W^{-}$ production at hadron colliders},''
  \href{http://dx.doi.org/10.1103/PhysRevD.53.1098}{{\em Phys. Rev.} {\bfseries
  D53} (1996) 1098--1123},
\href{http://arxiv.org/abs/hep-ph/9507336}{{\ttfamily arXiv:hep-ph/9507336
  [hep-ph]}}.

\bibitem{Dixon:1999di}
L.~J. Dixon, Z.~Kunszt, and A.~Signer, ``{Vector boson pair production in
  hadronic collisions at order $\alpha_s$ : Lepton correlations and anomalous
  couplings},'' \href{http://dx.doi.org/10.1103/PhysRevD.60.114037}{{\em Phys.
  Rev.} {\bfseries D60} (1999) 114037},
\href{http://arxiv.org/abs/hep-ph/9907305}{{\ttfamily arXiv:hep-ph/9907305
  [hep-ph]}}.

\bibitem{Chiesa:2018lcs}
M.~Chiesa, A.~Denner, and J.-N. Lang, ``{Anomalous triple-gauge-boson
  interactions in vector-boson pair production with RECOLA2},''
  \href{http://dx.doi.org/10.1140/epjc/s10052-018-5949-z}{{\em Eur. Phys. J.}
  {\bfseries C78} no.~6, (2018) 467},
\href{http://arxiv.org/abs/1804.01477}{{\ttfamily arXiv:1804.01477 [hep-ph]}}.

\bibitem{Campanario:2014lza}
F.~Campanario, R.~Roth, and D.~Zeppenfeld, ``{QCD radiation in $WH$ and $WZ$
  production and anomalous coupling measurements},''
  \href{http://dx.doi.org/10.1103/PhysRevD.91.054039}{{\em Phys. Rev.}
  {\bfseries D91} (2015) 054039},
\href{http://arxiv.org/abs/1410.4840}{{\ttfamily arXiv:1410.4840 [hep-ph]}}.

\bibitem{Granata:2017iod}
F.~Granata, J.~M. Lindert, C.~Oleari, and S.~Pozzorini, ``{NLO QCD+EW
  predictions for HV and HV +jet production including parton-shower effects},''
  \href{http://dx.doi.org/10.1007/JHEP09(2017)012}{{\em JHEP} {\bfseries 09}
  (2017) 012},
\href{http://arxiv.org/abs/1706.03522}{{\ttfamily arXiv:1706.03522 [hep-ph]}}.

\bibitem{Azatov:2019xxn}
A.~Azatov, D.~Barducci, and E.~Venturini, ``{Precision diboson measurements at
  hadron colliders},'' \href{http://dx.doi.org/10.1007/JHEP04(2019)075}{{\em
  JHEP} {\bfseries 04} (2019) 075},
\href{http://arxiv.org/abs/1901.04821}{{\ttfamily arXiv:1901.04821 [hep-ph]}}.

\bibitem{Contino:2016jqw}
R.~Contino, A.~Falkowski, F.~Goertz, C.~Grojean, and F.~Riva, ``{On the
  Validity of the Effective Field Theory Approach to SM Precision Tests},''
  \href{http://dx.doi.org/10.1007/JHEP07(2016)144}{{\em JHEP} {\bfseries 07}
  (2016) 144},
\href{http://arxiv.org/abs/1604.06444}{{\ttfamily arXiv:1604.06444 [hep-ph]}}.

\bibitem{Hays:2018zze}
C.~Hays, A.~Martin, V.~Sanz, and J.~Setford, ``{On the impact of
  dimension-eight SMEFT operators on Higgs measurements},''
  \href{http://dx.doi.org/10.1007/JHEP02(2019)123}{{\em JHEP} {\bfseries 02}
  (2019) 123},
\href{http://arxiv.org/abs/1808.00442}{{\ttfamily arXiv:1808.00442 [hep-ph]}}.

\bibitem{Alte:2018xgc}
S.~Alte, M.~Konig, and W.~Shepherd, ``{Consistent Searches for SMEFT Effects in
  Non-Resonant Dilepton Events},''
  \href{http://dx.doi.org/10.1007/JHEP07(2019)144}{{\em JHEP} {\bfseries 07}
  (2019) 144},
\href{http://arxiv.org/abs/1812.07575}{{\ttfamily arXiv:1812.07575 [hep-ph]}}.

\bibitem{Falkowski:2015jaa}
A.~Falkowski, M.~Gonzalez-Alonso, A.~Greljo, and D.~Marzocca, ``{Global
  constraints on anomalous triple gauge couplings in effective field theory
  approach},'' \href{http://dx.doi.org/10.1103/PhysRevLett.116.011801}{{\em
  Phys. Rev. Lett.} {\bfseries 116} no.~1, (2016) 011801},
\href{http://arxiv.org/abs/1508.00581}{{\ttfamily arXiv:1508.00581 [hep-ph]}}.

\bibitem{Franceschini:2017xkh}
R.~Franceschini, G.~Panico, A.~Pomarol, F.~Riva, and A.~Wulzer, ``{Electroweak
  Precision Tests in High-Energy Diboson Processes},''
  \href{http://dx.doi.org/10.1007/JHEP02(2018)111}{{\em JHEP} {\bfseries 02}
  (2018) 111},
\href{http://arxiv.org/abs/1712.01310}{{\ttfamily arXiv:1712.01310 [hep-ph]}}.

\bibitem{Liu:2019vid}
D.~Liu and L.-T. Wang, ``{Prospects for precision measurement of diboson
  processes in the semileptonic decay channel in future LHC runs},''
  \href{http://dx.doi.org/10.1103/PhysRevD.99.055001}{{\em Phys. Rev.}
  {\bfseries D99} no.~5, (2019) 055001},
\href{http://arxiv.org/abs/1804.08688}{{\ttfamily arXiv:1804.08688 [hep-ph]}}.

\bibitem{Falkowski:2015wza}
A.~Falkowski, B.~Fuks, K.~Mawatari, K.~Mimasu, F.~Riva, and V.~Sanz,
  ``{Rosetta: an operator basis translator for Standard Model effective field
  theory},'' \href{http://dx.doi.org/10.1140/epjc/s10052-015-3806-x}{{\em Eur.
  Phys. J.} {\bfseries C75} no.~12, (2015) 583},
\href{http://arxiv.org/abs/1508.05895}{{\ttfamily arXiv:1508.05895 [hep-ph]}}.

\bibitem{Dedes:2017zog}
A.~Dedes, W.~Materkowska, M.~Paraskevas, J.~Rosiek, and K.~Suxho, ``{Feynman
  rules for the Standard Model Effective Field Theory in R$_{\xi}$-gauges},''
  \href{http://dx.doi.org/10.1007/JHEP06(2017)143}{{\em JHEP} {\bfseries 06}
  (2017) 143},
\href{http://arxiv.org/abs/1704.03888}{{\ttfamily arXiv:1704.03888 [hep-ph]}}.

\bibitem{Brivio:2017bnu}
I.~Brivio and M.~Trott, ``{Scheming in the SMEFT... and a reparameterization
  invariance!},'' \href{http://dx.doi.org/10.1007/JHEP05(2018)136,
  10.1007/JHEP07(2017)148}{{\em JHEP} {\bfseries 07} (2017) 148},
  \href{http://arxiv.org/abs/1701.06424}{{\ttfamily arXiv:1701.06424
  [hep-ph]}}.
[Addendum: JHEP05,136(2018)].

\bibitem{Jenkins:2017jig}
E.~E. Jenkins, A.~V. Manohar, and P.~Stoffer, ``{Low-Energy Effective Field
  Theory below the Electroweak Scale: Operators and Matching},''
  \href{http://dx.doi.org/10.1007/JHEP03(2018)016}{{\em JHEP} {\bfseries 03}
  (2018) 016},
\href{http://arxiv.org/abs/1709.04486}{{\ttfamily arXiv:1709.04486 [hep-ph]}}.

\bibitem{Zhang:2016zsp}
Z.~Zhang, ``{Time to Go Beyond Triple-Gauge-Boson-Coupling Interpretation of
  $W$ Pair Production},''
  \href{http://dx.doi.org/10.1103/PhysRevLett.118.011803}{{\em Phys. Rev.
  Lett.} {\bfseries 118} no.~1, (2017) 011803},
\href{http://arxiv.org/abs/1610.01618}{{\ttfamily arXiv:1610.01618 [hep-ph]}}.

\bibitem{Berthier:2015oma}
L.~Berthier and M.~Trott, ``{Towards consistent Electroweak Precision Data
  constraints in the SMEFT},''
  \href{http://dx.doi.org/10.1007/JHEP05(2015)024}{{\em JHEP} {\bfseries 05}
  (2015) 024},
\href{http://arxiv.org/abs/1502.02570}{{\ttfamily arXiv:1502.02570 [hep-ph]}}.

\bibitem{Gupta:2014rxa}
R.~S. Gupta, A.~Pomarol, and F.~Riva, ``{BSM Primary Effects},''
  \href{http://dx.doi.org/10.1103/PhysRevD.91.035001}{{\em Phys. Rev.}
  {\bfseries D91} no.~3, (2015) 035001},
\href{http://arxiv.org/abs/1405.0181}{{\ttfamily arXiv:1405.0181 [hep-ph]}}.

\bibitem{Falkowski:2015fla}
A.~Falkowski, ``{Effective field theory approach to LHC Higgs data},''
  \href{http://dx.doi.org/10.1007/s12043-016-1251-5}{{\em Pramana} {\bfseries
  87} no.~3, (2016) 39},
\href{http://arxiv.org/abs/1505.00046}{{\ttfamily arXiv:1505.00046 [hep-ph]}}.

\bibitem{Tanabashi:2018oca}
{\bfseries Particle Data Group} Collaboration, M.~Tanabashi {\em et~al.},
  ``{Review of Particle Physics},''
\href{http://dx.doi.org/10.1103/PhysRevD.98.030001}{{\em Phys. Rev.} {\bfseries
  D98} no.~3, (2018) 030001}.

\bibitem{Alioli:2017ces}
S.~Alioli, V.~Cirigliano, W.~Dekens, J.~de~Vries, and E.~Mereghetti,
  ``{Right-handed charged currents in the era of the Large Hadron Collider},''
  \href{http://dx.doi.org/10.1007/JHEP05(2017)086}{{\em JHEP} {\bfseries 05}
  (2017) 086},
\href{http://arxiv.org/abs/1703.04751}{{\ttfamily arXiv:1703.04751 [hep-ph]}}.

\bibitem{Baur:1994ia}
U.~Baur, T.~Han, and J.~Ohnemus, ``{Amplitude zeros in $W^\pm Z$ production},''
  \href{http://dx.doi.org/10.1103/PhysRevLett.72.3941}{{\em Phys. Rev. Lett.}
  {\bfseries 72} (1994) 3941--3944},
\href{http://arxiv.org/abs/hep-ph/9403248}{{\ttfamily arXiv:hep-ph/9403248
  [hep-ph]}}.

\bibitem{Panico:2017frx}
G.~Panico, F.~Riva, and A.~Wulzer, ``{Diboson Interference Resurrection},''
  \href{http://dx.doi.org/10.1016/j.physletb.2017.11.068}{{\em Phys. Lett.}
  {\bfseries B776} (2018) 473--480},
\href{http://arxiv.org/abs/1708.07823}{{\ttfamily arXiv:1708.07823 [hep-ph]}}.

\bibitem{Brehmer:2019gmn}
J.~Brehmer, S.~Dawson, S.~Homiller, F.~Kling, and T.~Plehn, ``{Benchmarking
  simplified template cross sections in $WH$ production},''
  \href{http://dx.doi.org/10.1007/JHEP11(2019)034}{{\em JHEP} {\bfseries 11}
  (2019) 034},
\href{http://arxiv.org/abs/1908.06980}{{\ttfamily arXiv:1908.06980 [hep-ph]}}.

\bibitem{Aaboud:2019gxl}
{\bfseries ATLAS} Collaboration, M.~Aaboud {\em et~al.}, ``{Measurement of
  $W^{\pm}Z$ production cross sections and gauge boson polarisation in $pp$
  collisions at $\sqrt{s} = 13$ TeV with the ATLAS detector},''
  \href{http://dx.doi.org/10.1140/epjc/s10052-019-7027-6}{{\em Eur. Phys. J.}
  {\bfseries C79} no.~6, (2019) 535},
\href{http://arxiv.org/abs/1902.05759}{{\ttfamily arXiv:1902.05759 [hep-ex]}}.

\bibitem{Bern:2011ie}
Z.~Bern {\em et~al.}, ``{Left-Handed W Bosons at the LHC},''
  \href{http://dx.doi.org/10.1103/PhysRevD.84.034008}{{\em Phys. Rev.}
  {\bfseries D84} (2011) 034008},
\href{http://arxiv.org/abs/1103.5445}{{\ttfamily arXiv:1103.5445 [hep-ph]}}.

\bibitem{Falkowski:2016cxu}
A.~Falkowski, M.~Gonzalez-Alonso, A.~Greljo, D.~Marzocca, and M.~Son,
  ``{Anomalous Triple Gauge Couplings in the Effective Field Theory Approach at
  the LHC},'' \href{http://dx.doi.org/10.1007/JHEP02(2017)115}{{\em JHEP}
  {\bfseries 02} (2017) 115},
\href{http://arxiv.org/abs/1609.06312}{{\ttfamily arXiv:1609.06312 [hep-ph]}}.

\bibitem{Dixon:1993xd}
L.~J. Dixon and Y.~Shadmi, ``{Testing gluon selfinteractions in three jet
  events at hadron colliders},''
  \href{http://dx.doi.org/10.1016/0550-3213(94)90563-0,
  10.1016/0550-3213(95)00450-7}{{\em Nucl. Phys.} {\bfseries B423} (1994)
  3--32}, \href{http://arxiv.org/abs/hep-ph/9312363}{{\ttfamily
  arXiv:hep-ph/9312363 [hep-ph]}}.
[Erratum: Nucl. Phys.B452,724(1995)].

\bibitem{Azatov:2017kzw}
A.~Azatov, J.~Elias-Miro, Y.~Reyimuaji, and E.~Venturini, ``{Novel measurements
  of anomalous triple gauge couplings for the LHC},''
  \href{http://dx.doi.org/10.1007/JHEP10(2017)027}{{\em JHEP} {\bfseries 10}
  (2017) 027},
\href{http://arxiv.org/abs/1707.08060}{{\ttfamily arXiv:1707.08060 [hep-ph]}}.

\bibitem{Aaboud:2019nan}
{\bfseries ATLAS} Collaboration, M.~Aaboud {\em et~al.}, ``{Measurement of VH,
  $ \mathrm{H}\to \mathrm{b}\overline{\mathrm{b}} $ production as a function of
  the vector-boson transverse momentum in 13 TeV pp collisions with the ATLAS
  detector},'' \href{http://dx.doi.org/10.1007/JHEP05(2019)141}{{\em JHEP}
  {\bfseries 05} (2019) 141},
\href{http://arxiv.org/abs/1903.04618}{{\ttfamily arXiv:1903.04618 [hep-ex]}}.

\bibitem{Aad:2016wpd}
{\bfseries ATLAS} Collaboration, G.~Aad {\em et~al.}, ``{Measurement of total
  and differential $W^+W^-$ production cross sections in proton-proton
  collisions at $\sqrt{s}=$ 8 TeV with the ATLAS detector and limits on
  anomalous triple-gauge-boson couplings},''
  \href{http://dx.doi.org/10.1007/JHEP09(2016)029}{{\em JHEP} {\bfseries 09}
  (2016) 029},
\href{http://arxiv.org/abs/1603.01702}{{\ttfamily arXiv:1603.01702 [hep-ex]}}.

\bibitem{Aaboud:2019nkz}
{\bfseries ATLAS} Collaboration, M.~Aaboud {\em et~al.}, ``{Measurement of
  fiducial and differential $W^+W^-$ production cross-sections at $\sqrt{s}=13$
  TeV with the ATLAS detector},''
  \href{http://dx.doi.org/10.1140/epjc/s10052-019-7371-6}{{\em Eur. Phys. J.}
  {\bfseries C79} no.~10, (2019) 884},
\href{http://arxiv.org/abs/1905.04242}{{\ttfamily arXiv:1905.04242 [hep-ex]}}.

\bibitem{Aad:2016ett}
{\bfseries ATLAS} Collaboration, G.~Aad {\em et~al.}, ``{Measurements of $W^\pm
  Z$ production cross sections in $pp$ collisions at $\sqrt{s} = 8$ TeV with
  the ATLAS detector and limits on anomalous gauge boson self-couplings},''
  \href{http://dx.doi.org/10.1103/PhysRevD.93.092004}{{\em Phys. Rev.}
  {\bfseries D93} no.~9, (2016) 092004},
\href{http://arxiv.org/abs/1603.02151}{{\ttfamily arXiv:1603.02151 [hep-ex]}}.

\bibitem{Khachatryan:2016poo}
{\bfseries CMS} Collaboration, V.~Khachatryan {\em et~al.}, ``{Measurement of
  the WZ production cross section in pp collisions at $\sqrt{s} = 7$ and 8
  $\,\text{TeV}$ and search for anomalous triple gauge couplings at $\sqrt{s} =
  8\,\text{TeV} $},''
  \href{http://dx.doi.org/10.1140/epjc/s10052-017-4730-z}{{\em Eur. Phys. J.}
  {\bfseries C77} no.~4, (2017) 236},
\href{http://arxiv.org/abs/1609.05721}{{\ttfamily arXiv:1609.05721 [hep-ex]}}.

\bibitem{Sirunyan:2019bez}
{\bfseries CMS} Collaboration, A.~M. Sirunyan {\em et~al.}, ``{Measurements of
  the pp $\to$ WZ inclusive and differential production cross section and
  constraints on charged anomalous triple gauge couplings at $\sqrt{s} =$ 13
  TeV},'' \href{http://dx.doi.org/10.1007/JHEP04(2019)122}{{\em JHEP}
  {\bfseries 04} (2019) 122},
\href{http://arxiv.org/abs/1901.03428}{{\ttfamily arXiv:1901.03428 [hep-ex]}}.

\bibitem{Falkowski:2014tna}
A.~Falkowski and F.~Riva, ``{Model-independent precision constraints on
  dimension-6 operators},''
  \href{http://dx.doi.org/10.1007/JHEP02(2015)039}{{\em JHEP} {\bfseries 02}
  (2015) 039},
\href{http://arxiv.org/abs/1411.0669}{{\ttfamily arXiv:1411.0669 [hep-ph]}}.

\bibitem{Dawson:2019clf}
S.~Dawson and P.~P. Giardino, ``{Electroweak and QCD corrections to $Z$ and $W$
  pole observables in the standard model EFT},''
  \href{http://dx.doi.org/10.1103/PhysRevD.101.013001}{{\em Phys. Rev.}
  {\bfseries D101} no.~1, (2020) 013001},
\href{http://arxiv.org/abs/1909.02000}{{\ttfamily arXiv:1909.02000 [hep-ph]}}.

\end{thebibliography}\endgroup

\end{document}